# Prompt fission neutron spectra of $^{233}$U($n,F$)

V. M. Maslov[1] (orcid :0000-0001-7247-2443)

*220025 Minsk, Byelorussia*

**Abstract:** Observed prompt fission neutron spectra (PFNS) for incident neutron energies $E_n$ from thermal $E_{th}$ up to 20 MeV are employed to disentangle pre- and post-fission components of prompt fission neutrons. Simultaneous analysis of measured and calculated data for $^{233}$U($n,F$), $^{235}$U($n,F$) and $^{239}$Pu($n,F$) maintains stronger justification for the predicted pre- and post-fission components PFNS of $^{233}$U($n,F$). For the $^{233}$U($n,F$) reaction the reliable measured PFNS data are available at $E_n \sim E_{th}$ only. Pre–fission neutron spectra influence the partitioning of fission energy between excitation energy and total kinetic energy of fission fragments. For the reactions $^{233}$U($n,F$) and $^{235}$U($n,F$) the shape of prompt fission neutron spectra (PFNS) depend on the fissility of composite and residual nuclides in a similar way. The correlation of these peculiarities with the emissive fission contributions ($n,xnf$) to the observed fission cross section and the competition of the reactions ($n,n\gamma$) and $(n,xn)^{1,..x}$ is established. Calculated exclusive neutron spectra $(n,xnf)^{1,..x}$ are consistent with the fission cross sections of $^{235}$U($n,F$), $^{234}$U($n,F$), $^{233}$U($n,F$) and $^{232}$U($n,F$) reactions, as well as available neutron emissive spectra of $^{235}$U($n,xn$) at ~14 MeV. Initial model parameters for the PFNS of $^{233}$U($n,F$) reaction calculation are fixed by the description of PFNS of $^{233}$U($n_{th},F$). The $^{233}$U$(n,xnf)^{1,..x}$ exclusive pre–fission neutron spectra and exclusive neutron spectra of $^{233}$U$(n,xn)^{1,..x}$ reactions, total kinetic energy TKE of fission fragments and products, partials of average prompt fission neutron number and observed PFNS of $^{233}$U($n,F$) are predicted. Prompt fission neutron spectra of $^{233}$U($n,F$) are harder than those of $^{235}$U($n,F$) PFNS, but softer than those of the $^{239}$Pu($n,F$) reaction. Difference of average energies of PFNS $\langle E \rangle$ of $^{233}$U($n,F$) and $^{235}$U($n,F$) amounts to 1~3 %. At incident neutron energies higher than ($n,2nf$) reaction threshold, the observed PFNS seem rather similar, though the partial contributions of $^{233}$U($n,xnf$) and $^{235}$U($n,xnf$) reactions to the observed PFNS are quite different from each other. The similarity of dips in $\langle E \rangle$ of $^{233}$U($n,F$) and $^{235}$U($n,F$) PFNS due to $(n,xnf)^{1,..x}$ neutrons (the depth and obvious asymmetry relative to the lowest value) is confirmed by the $^{233}$U($n,F$) and $^{235}$U($n,F$) PF fission fragments angular asymmetry with respect to the incident neutron beam momentum. The similarity of dips in average energies $\langle E \rangle$ of $^{233}$U($n,F$) and $^{235}$U($n,F$) PFNS (the depth and obvious dip' asymmetry relative to the largest amplitude) is supported by the fission fragments angular asymmetry with respect to the incident neutron momentum.

## 1. INTRODUCTION

Fissile nuclides $^{233}$U build up in a breeder or hybrid reactors via reaction chains

[1]E–mail: mvm2386@yandex.ru

$$^{232}Th+n \to {}^{233}Th(22.3\ min) \to {}^{233}Pa(27\ d) \to {}^{233}U.$$

Nuclear data on $^{233}U$+n interaction, with the exception of fission cross section, are scarce, especially as regards differential prompt fission neutron spectra $S(\varepsilon, E_n)$ of $^{233}U(n,F)$ reaction in the excitation energy range where exclusive pre-fission neutron spectra $^{233}U(n,xnf)^{1,..x}$ are mixed with the prompt fission neutron spectra emitted by the fission fragments. Prompt fission neutron spectra $S(\varepsilon, E_n)$ measured at $E_n$~14.3 MeV [1] remained the only available open data before long. Prompt fission neutron spectra in [1] were registered in a rather narrow energy range of ε~0.4÷5 MeV, however the excess of soft pre-fission neutrons in $S(\varepsilon, E_n)$ at ε~0.4÷2 MeV energy range was observed distinctly. Afterwards measured data on PFNS were described with a straightforward superposition of Weisscopf' [2] and Watt' [3] type distributions:

$$S(\varepsilon, E_n) = (1-\beta)\varepsilon \exp\left(-\frac{\varepsilon}{T}\right) + \beta \exp\left(-\frac{\varepsilon}{T_f}\right)\frac{sh(2\sqrt{\omega\varepsilon})}{T_f}, \qquad (1)$$

here β=0.75 – value of lumped contribution of post-fission neutrons, emitted from fission fragments, average temperature of residual nuclide $T$=0.55 MeV, fission fragments temperature $T_f$ = 1.2 MeV, parameter ω=0.5 MeV. Since Eq. (1) looks over-simplified and the energy range of ε~0.4÷5 MeV is rather narrow, relative contributions of pre- and post-fission neutrons in [1] disagree with newer research data [4–6]. However, the shape of the calculated PFNS in [1] looks roughly similar to that elaborated in [4–6]. Available measured data at $E_n$~$E_{th}$ [7] and $E_n$~0.55 MeV [8], and data [1] as well, were abandoned in a numerous versions of ENDF/B, JEFF and JENDL evaluated data libraries. The data of recent PFNS measurements of $^{233}U(n_{th},f)$ [9], $^{235}U(n_{th},f)$ and $^{239}Pu(n_{th},f)$ [9, 10] are quite different as compared with the calculations of [4–6]. Lumping data of [8, 9] within spline fitting procedure [11] would change resulting PFNS shapes of $^{233}U(n_{th},f)$, $^{235}U(n_{th},f)$ and $^{239}Pu(n_{th},f)$ drastically (see Fig. 1 and Fig. 2).

Differential PFNS data for the target nuclides $^{235}U$ and $^{239}Pu$ available for the incident neutron energies $E_n$~1.5÷20 MeV and outgoing neutron energies ε~0.01÷10 MeV [12–14] help to extract a number of peculiarities as dependent upon the target nuclide neutron-induced fission cross section. Strong variations of the average energies of prompt fission neutrons in the vicinity of (*n,xnf*) reaction thresholds were observed in [12–14]. Average energies of prompt fission neutrons $\langle E \rangle$ are rough signature of PFNS, however it was established in the measurements [12–14] that the relative amplitude of $\langle E \rangle$ variation in case of $^{239}Pu(n,F)$ reaction is much weaker than in case of $^{235}U(n,F)$ PFNS average energy $\langle E \rangle$. That peculiarity is strongly correlated with the influence of (*n,xnf*) fission channels on fission observables, when emission of prompt fission neutrons from the fragments is preceded by the pre-fission neutrons. In case of $^{233}U(n,F)$ reactions the variations of $\langle E \rangle$ were predicted in [5, 6, 11], as they are strictly correlated with the shape of calculated PFNS.

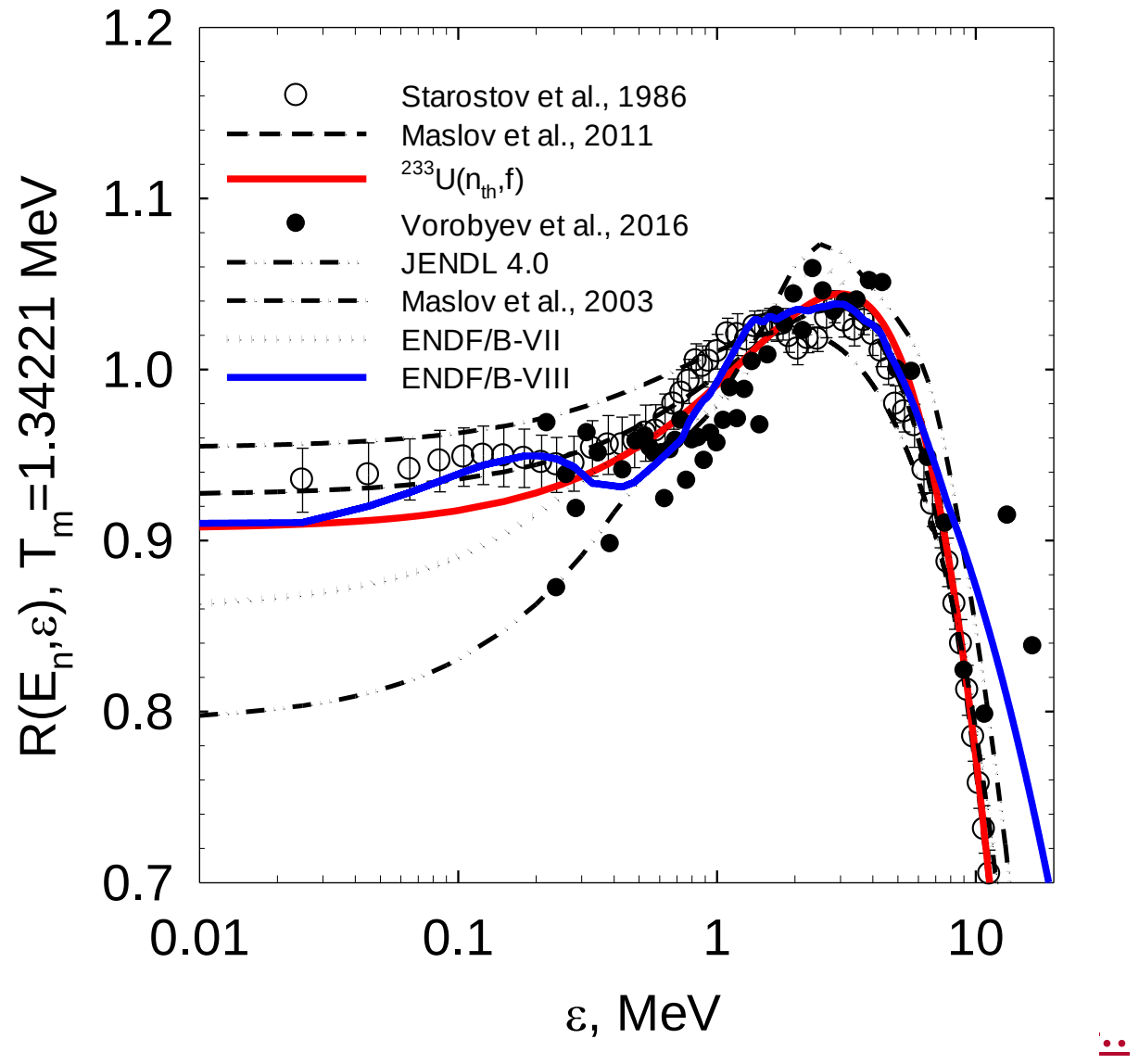

Fig.1. Prompt fission neutron spectra of $^{233}$U($n_{th}$,$f$) relative to the Maxwellian type distribution with $\langle E \rangle$ = 2,0564 MeV: the red solid line corresponds to $^{233}$U($n_{th}$,$f$), the black dash dotted line is data from [4, 5], the black dashed line is from [11], the black dotted line is from [15], the blue solid line is from [17], the double-dotted-dashed line is from [18], the white dots are from [7], the black dots are from [9].

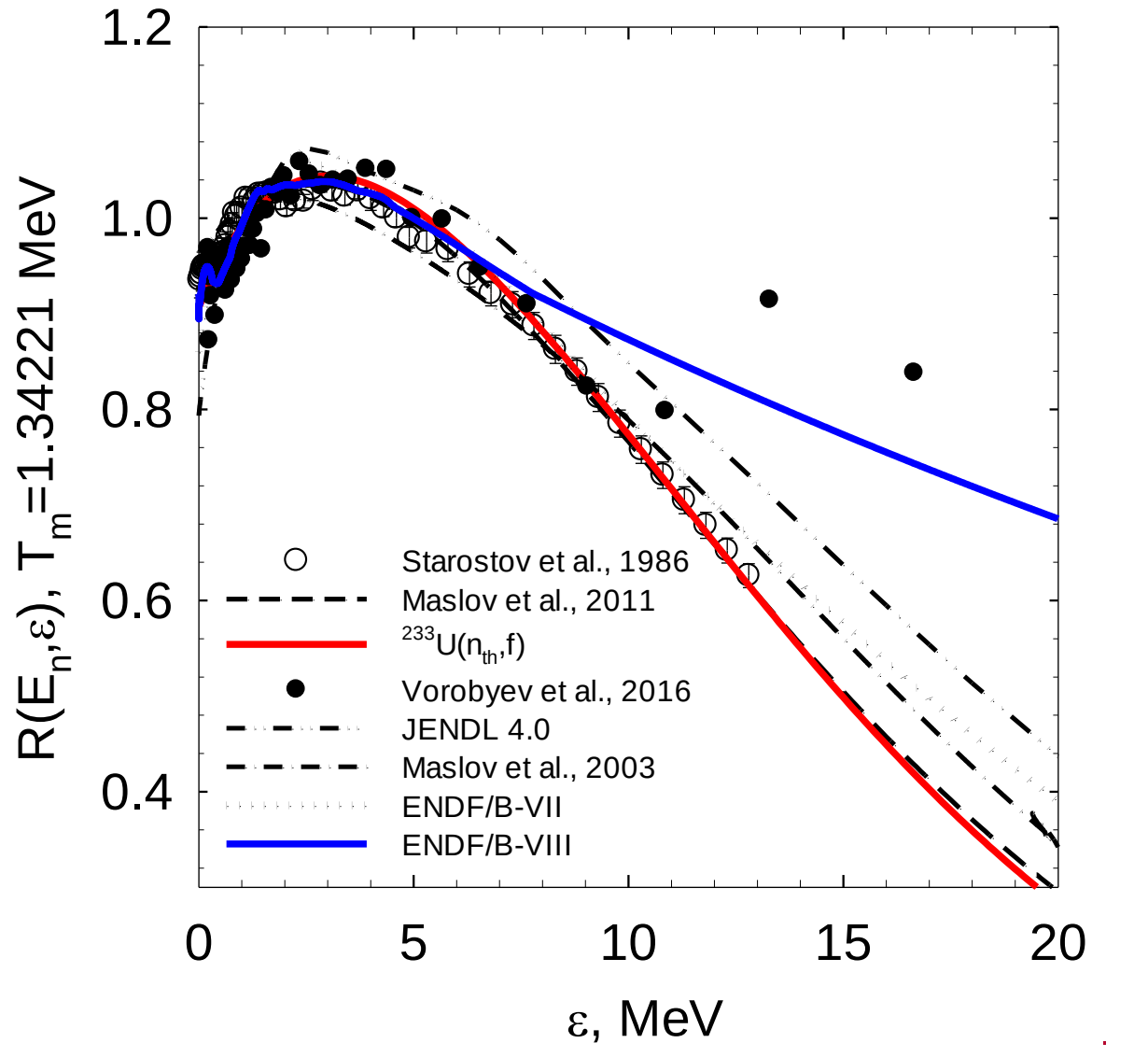

Fig.2. Same as shown in Fig. 1, but the neutron energy ε is linear scaled.

Variations of $\langle E \rangle$ for $^{233}$U($n,F$) of ENDF/B–VII [15] are due to the arbitrary variation of PFNS shape to reproduce some assumed dependence of $\langle E \rangle$($E_n$). The $^{233}$U+$n$ data file of ENDF/B–VII [15] is borrowed for the JEFF–3.3 data library [16] with some minor exceptions. In the version of ENDF/B–VIII.0 [17] they abruptly adopted JENDL–4.0 [18] calculations and waived off data file of ENDF/B–VII [15]. In the ROSFOND library [19] they use $^{233}$U data file of [4, 5]. In BROND library [20] only prompt fission neutron spectra of $^{233}$U [4, 5] are adopted, however that is rather controversial procedure, since the PFNS shape depends on the calculated $^{233}$U+$n$ interaction data.

We intend to define/predict PFNS of $^{233}$U($n,F$) reaction in the energy range $E_n$~ $E_{th}$÷20 MeV and extract pre- and post-fission components of prompt fission neutrons. For that purpose methods, proved by the $^{239}$Pu($n,F$) and $^{235}$U($n,F$) PFNS data analysis and prediction, are extensively used.

## 2. Prompt fission neutrons

Measured PFNS data of $^{233}$U($n, F$) usually add more controversy to the data base first provoked by the data of [7]. As follows from the analysis of $^{233}$U($n_{th},f$) PFNS data [9], in the outgoing neutron energy range of 0.02<ε<5 MeV, they further support the calculations adopted for ENDF/B–VII [15] and JENDL–4.0 [18], while both calculated PFNS of later versions of the evaluations [15, 18] strongly disagree with the data [7]. Data [7] are presented on figures 1, 2 and 3 as a spline approximation [11], which summons empirical features of the simultaneous analysis [11] of $^{233}$U($n_{th},f$), $^{235}$U($n_{th},f$), $^{239}$Pu($n_{th},f$) and $^{252}$Cf($sf$) measured PFNS data. A number of controversies are further revealed. In the energy range of 5<ε<11 MeV data of [9] for $^{233}$U($n_{th},f$) support the evaluation of [11], which is based on the data [7] available in the energy range of 0.02<ε<9.3 MeV. The shape of $^{233}$U($n_{th},f$) PFNS of ENDF/B–VII [15] conforms with the evaluated PFNS of $^{235}$U($n_{th},f$) [9]. However, it is well known that $^{235}$U($n_{th},f$) PFNS of ENDF/B–VII [15] only poorly describes measured data [21] at $E_n$~0.5 MeV and virtually disagree with the shape of newest measured data [12–14] at $E_n$≳1.5 MeV. Moreover, ENDF/B–VIII [17], JEFF–3.3 [16] and JENDL–4.0 [18] libraries use at $E_n$~$E_{th}$ the JENDL–3.3 [18] evaluated data, which are assumed uncorrelated with the PFNS shape at incident neutron energies $E_n$>$E_{th}$.

The comparison of PFNS measured data [7–9, 12–14] for $^{233}$U($n, f$), $^{235}$U($n, f$) and $^{239}$Pu($n, f$) in the energy range $E_{th}$ <$E_n$< $E_{nnf}$ shows that enhanced soft, ε≲1 MeV, neutron yield is a common feature unlike measured PFNS data [9, 10] at $E_n$=$E_{th}$ (see Fig. 1 and Fig. 2). In [9, 10] PFNS of $^{233}$U($n_{th},f$), $^{235}$U($n,f$) and $^{239}$Pu($n,f$) were measured relative to the spontaneous fission neutron spectra of $^{252}$Cf($sf$). Afterwards various correction were applied to get absolute PFNS values, possibly a number of systematic errors/uncertainties may appear, while the uncorrected measured ratios of various $^{233}$U($n_{th},f$), $^{235}$U($n_{th},f$) and $^{239}$Pu($n_{th},f$) PFNS pairs might be quite sterile in that respect.

The ratios of measured PFNS data of $^{239}$Pu($n_{th},f$)/$^{233}$U($n_{th},f$) and $^{233}$U($n_{th},f$)/$^{235}$U($n_{th},f$) of [9, 10] and [7], contrary to the absolute PFNS of $^{233}$U($n_{th},f$), $^{235}$U($n_{th},f$) and $^{239}$Pu($n_{th},f$), quite agree with each other (see Fig. 3 and Fig. 4). Uncorrected absolute data of PFNS [7, 9, 10] are unavailable at the moment, though it would be preferable to compare their ratios. Figures 3 and 4 demonstrate that in the energy range 0.01≲ε ≲10 MeV the ratios of calculated PFNS of $^{239}$Pu($n,f$) and $^{235}$U($n,f$) at $E_n$~$E_{th}$ and $E_n$~0.5 MeV, only weakly depend on the incident neutron energy $E_n$. Calculated PFNS ratios of [4–6, 11, 24], as well as present calculation, at $E_n$~$E_{th}$ and $E_n$~0.5 MeV, almost coincide with the measured

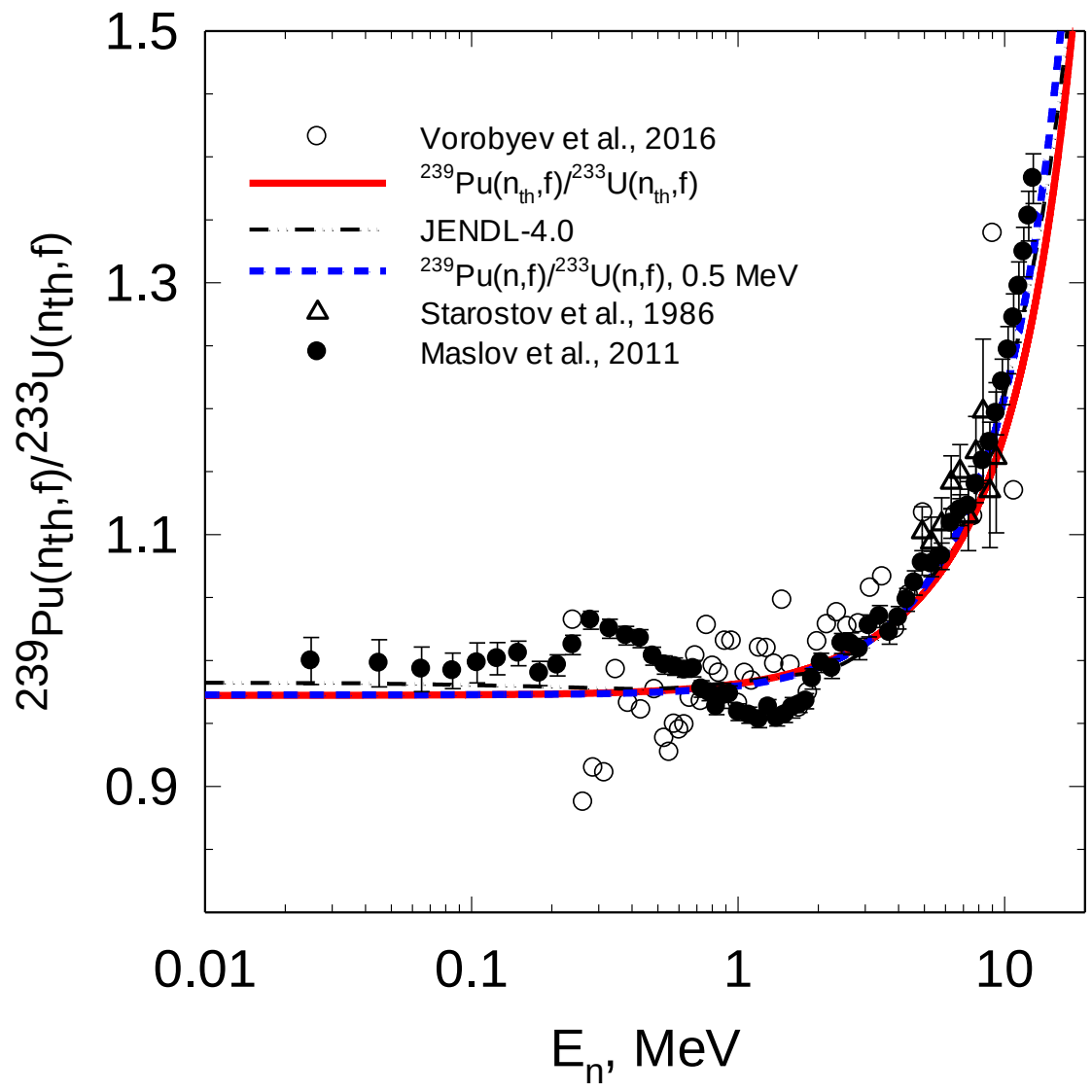

Fig.3. Ratio of PFNS of [239]Pu ($n_{th}$,*f*) and [233]U($n_{th}$,*f*) for thermal neutron-induced fission: the white triangles correspond to data from [7], the white dots are from [9], the black dots are from [11], the red solid line represents the current results, the double-dotted-dashed line is from [18], the blue dashed line represents the current results for the value of $E_n$ = 0.5 MeV.

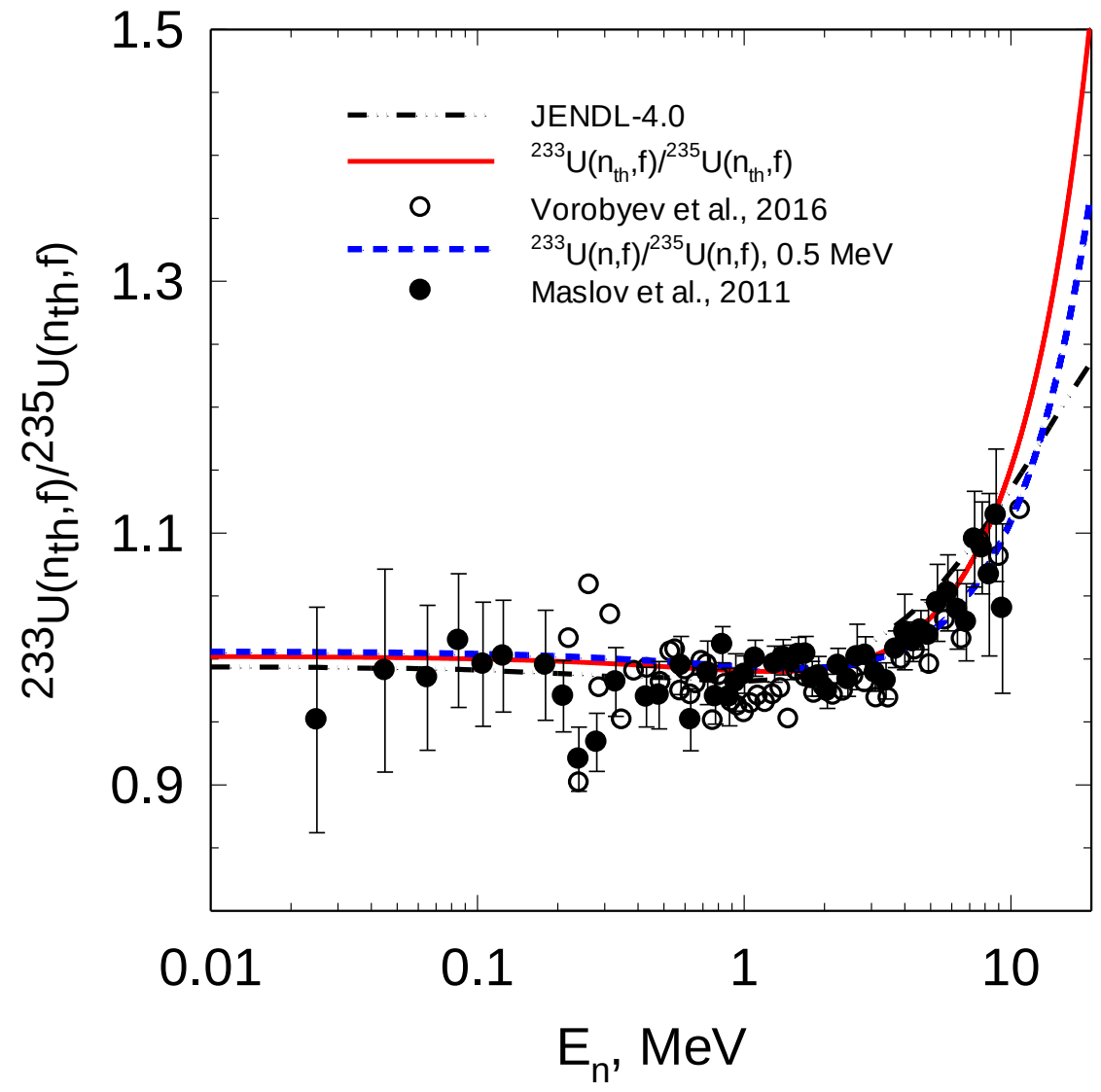

Fig.4. Ratio of PFNS of [233]U($n_{th}$,*f*) ($n_{th}$,*f*) and [235]U($n_{th}$,*f*) for thermal neutron-induced fission: the white dots corresponds to data from [9, 10], the black dots are from [11], the red solid line represents the current results, the double-dotted-dashed line is from [18], the blue dashed line represents the current results for the value of $E_n$ = 0.5 MeV.

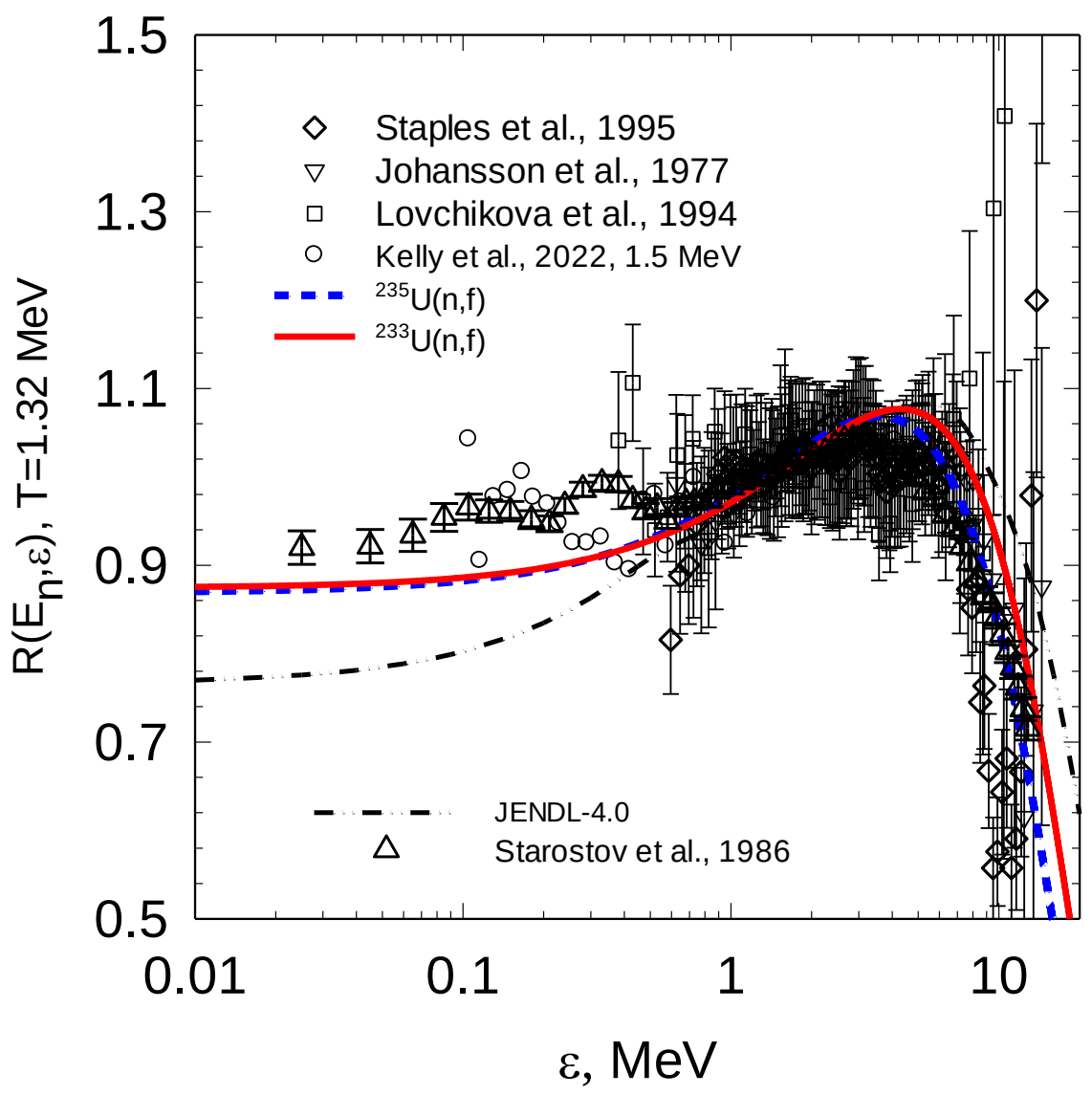

Fig.5 Prompt fission neutron spectra at $E_n$ = 0.5 MeV relative to the Maxwellian type distribution with $T$ =1.34175 MeV: the red solid line corresponds to $^{233}$U($n$, $F$), the blue dashed line corresponds to $^{235}$U($n$, $F$), the white triangles correspond to $^{233}$U($n_{th}F$) [7], the white dots correspond to $^{235}$U($n$, $F$) with the value of $E_n$ = 1—2 MeV [12], the down pointing triangles correspond to $^{235}$U($n$, $F$) [21], the diamonds correspond to $^{235}$U($n$, $F$) [22], the white rectangles correspond to $^{235}$U($n$,$F$) [23].

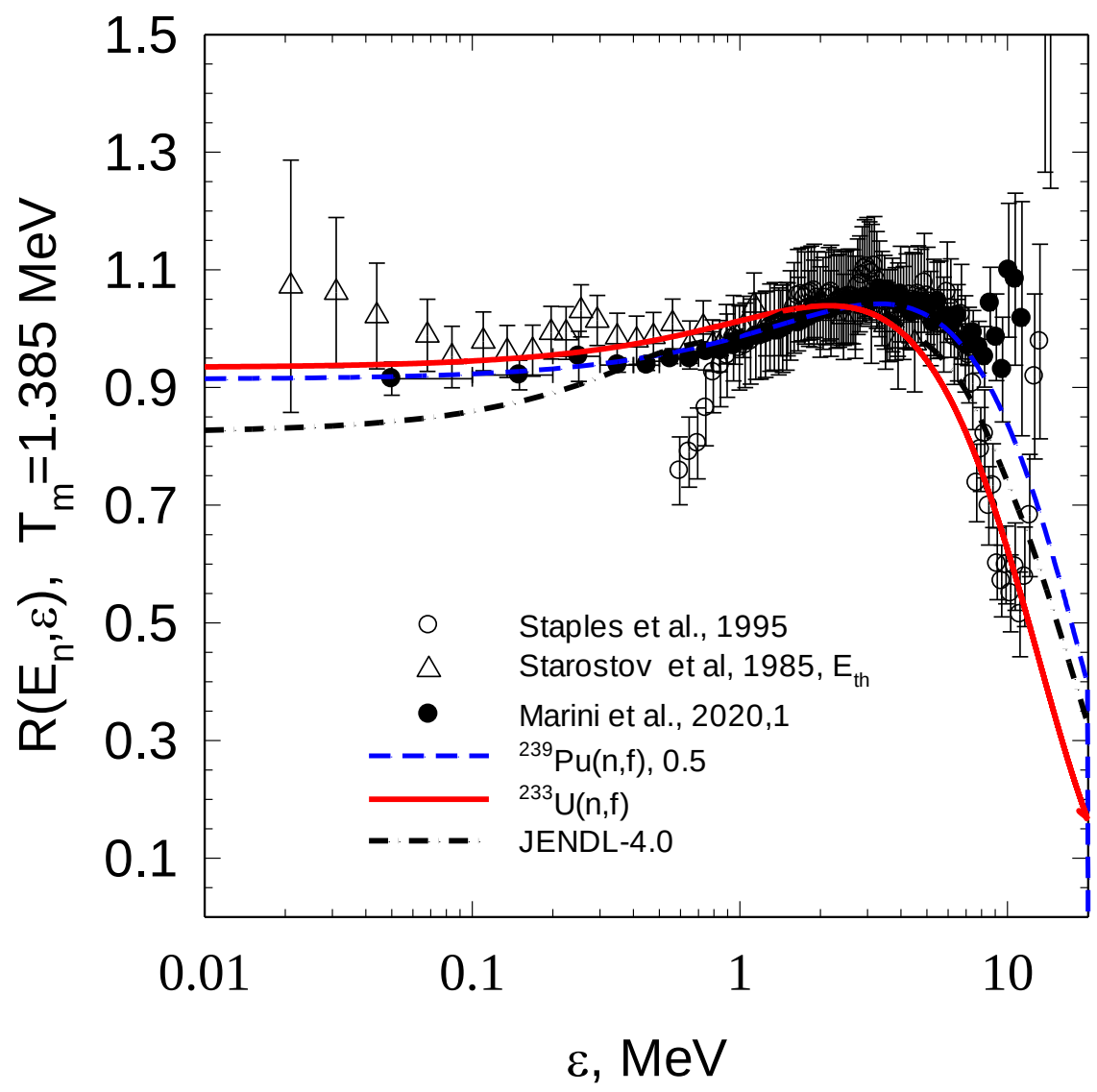

*

Fig.6. Prompt fission neutron spectra at $E_n$ = 0.5 MeV relative to the Maxwellian type distribution with $T$ =1.34175 MeV: the red solid line corresponds to $^{233}$U($n$,$F$), the dashdotted line corresponds to $^{233}$U($n$,$F$) [18], the blue dashed line corresponds to $^{239}$Pu($n$,$F$), the white triangles correspond to $^{239}$Pu($n$,$F$) [7], the white dots correspond to $^{239}$Pu($n$,$F$) [22] with the value of $E_n$ = 1−2 MeV, the black dots correspond to $^{239}$Pu($n$, $F$) [14].

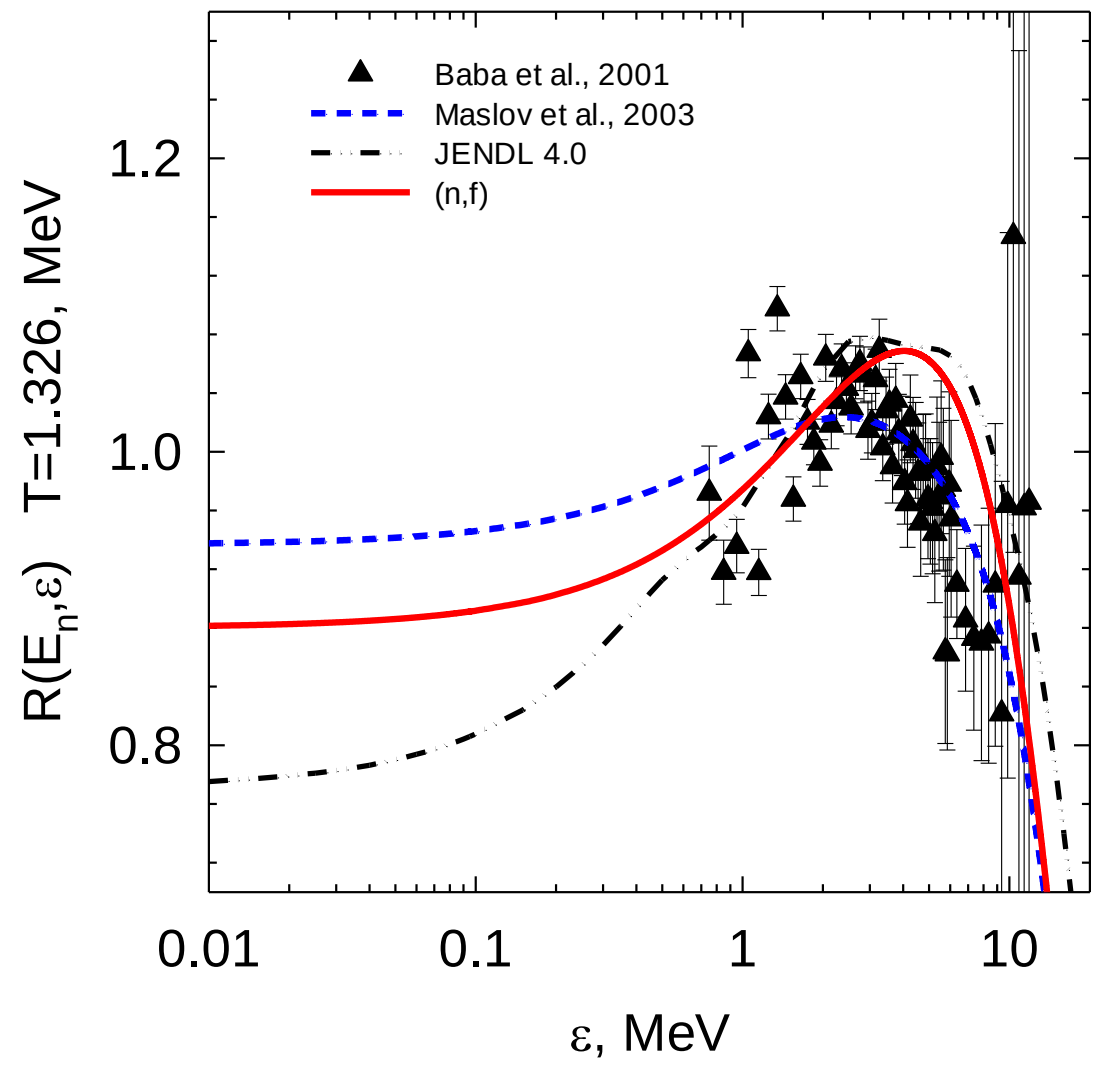

Fig.7. Prompt fission neutron spectra at $E_n$ = 0.5 MeV relative to the Maxwellian-type distribution with $T$ =1.326 MeV: the red solid line corresponds to $^{233}$U($n$,$F$), the double-dotted dashed line corresponds to $^{233}$U($n$,$F$) [18], the blue dashed line corresponds to data from [4, 5], and the black triangles correspond to data from [8].

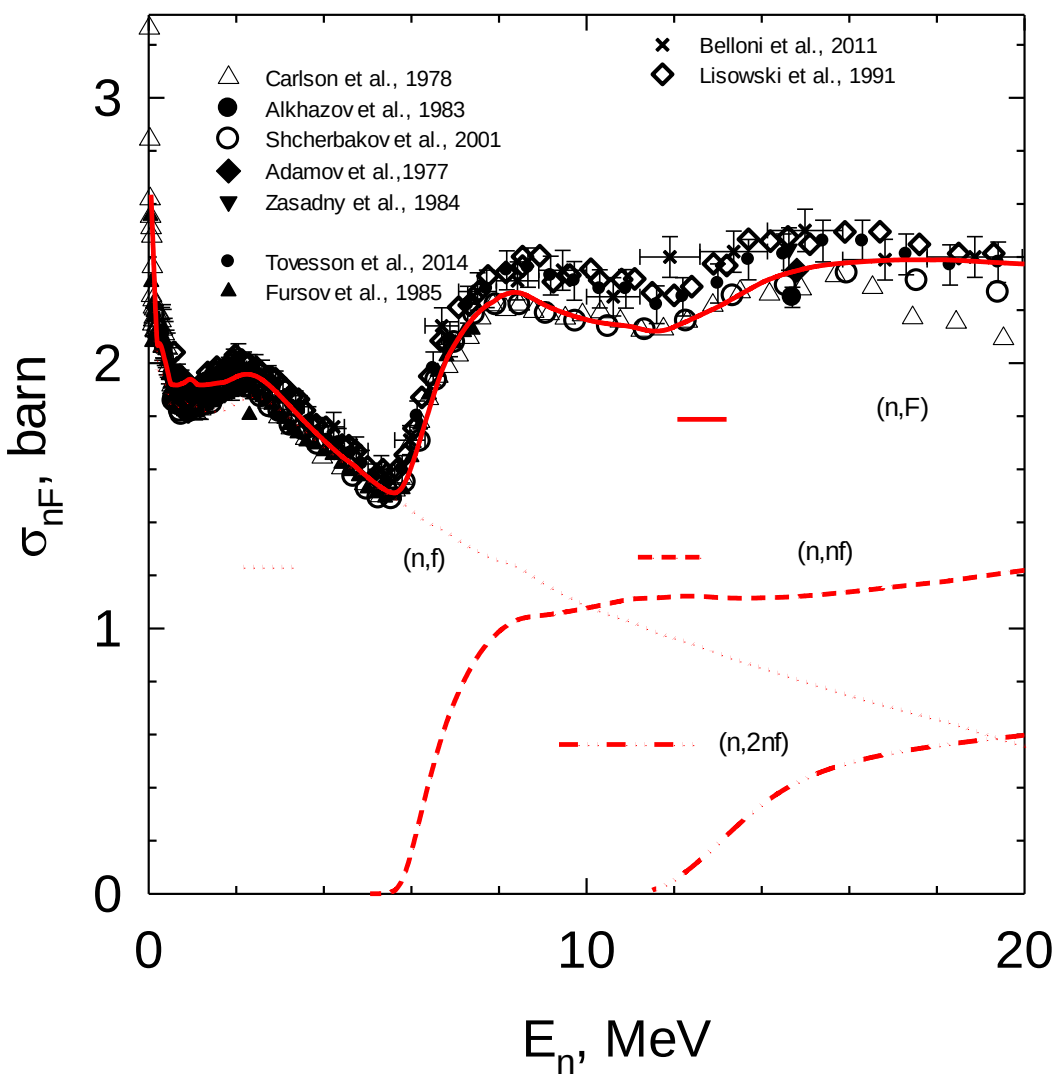

Fig.8. The data of $^{233}$U($n$,$F$) partial components according to the following references: the black triangles are from [44], the black diamonds are from [47], the black dots are from [48], the white dots are from [49], the white triangles are from [50], the black triangles are from [51], the white diamonds are from [52], the black small dots are from [53], the black crosses are from [54]. The red solid, dotted, dashed and double-dotted dashed lines correspond to ($n$,$F$), ($n$,$f$), ($n$,$nf$) and ($n$,$2nf$) reactions respectively.

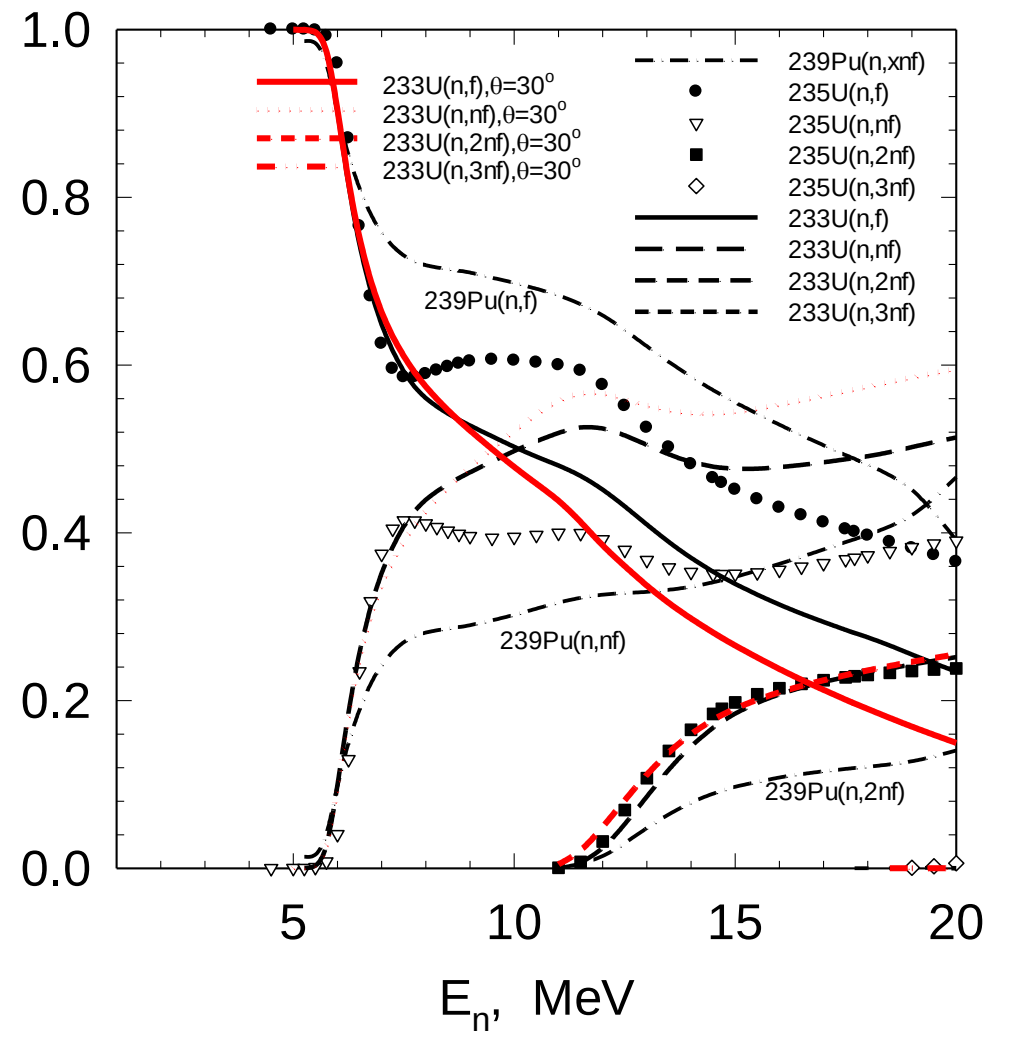

Fig.9. Ratios of partial components (*n, xnf*) to the observed fission cross section (*n,F*). The black dots correspond to $^{235}$U(*n,f*), the white triangles correspond to $^{235}$U(*n,nf*), the black rectangles correspond to $^{235}$U(*n,2nf*), the white diamonds correspond to $^{235}$U(*n,3nf*) reactions. The black solid line corresponds to $^{233}$U(*n,f*), the loosely dashed line corresponds to $^{233}$U(*n,nf*), the dashed line corresponds to $^{233}$U(*n,2nf*), the densely dashed line corresponds to $^{233}$U(*n,3nf*). The red solid line corresponds to $^{233}$U(*n,f*), the red dotted line corresponds to $^{233}$U(*n,nf*), the dashed line corresponds to $^{233}$U(*n,2nf*), the double-dotted dashed line corresponds to $^{233}$U(*n,3nf*) for $\theta$=30°. The dashdotted lines correspond to different $^{239}$Pu(*n, xnf*) reactions labeled in the figure.

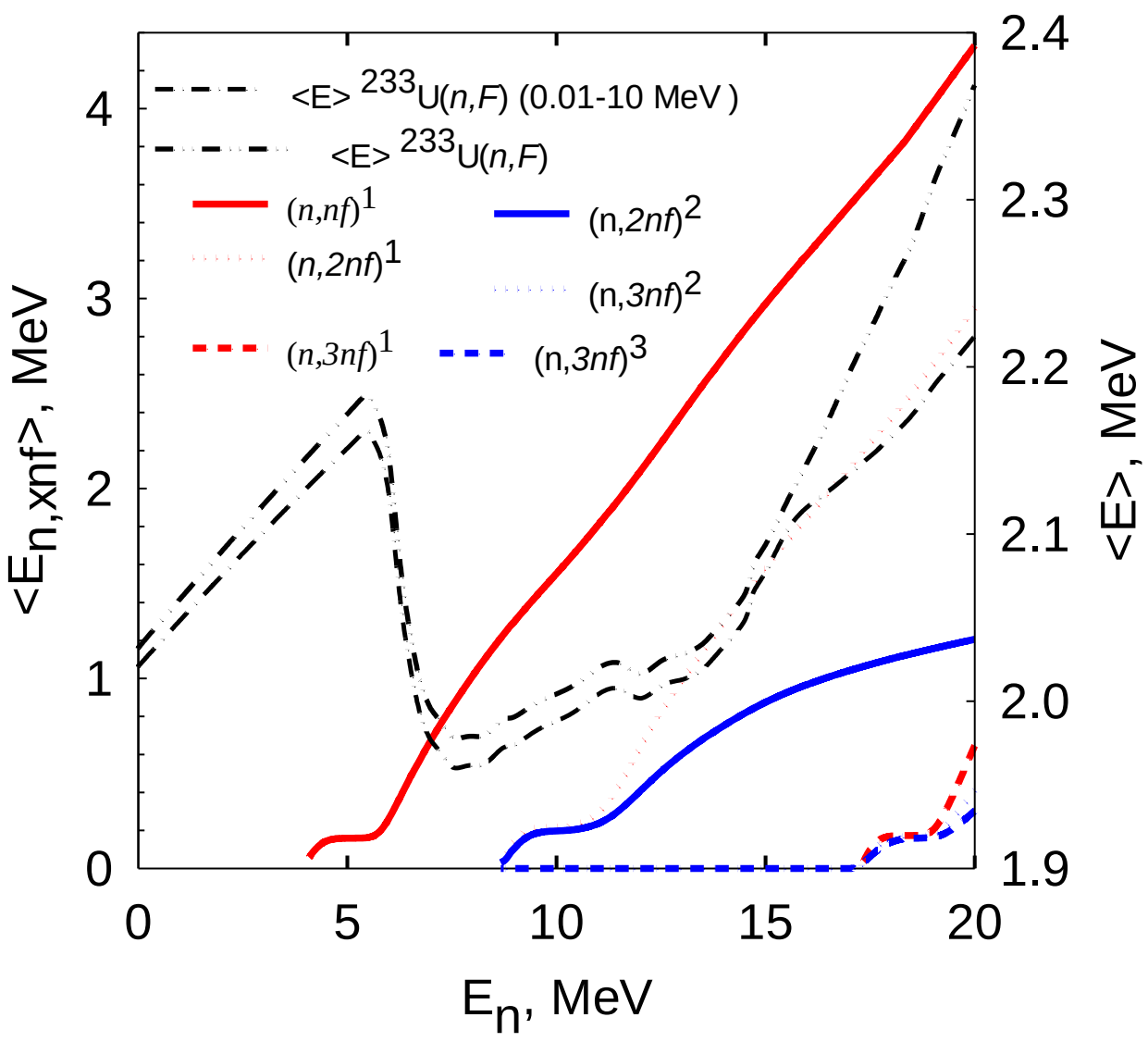

Fig.10. Average energy of exclusive pre-fission neutron spectra $^{233}$U(*n,xnf*): red solid, dotted and dashed lines correspond to $^{233}$U(*n,nf*)$^1$, $^{233}$U(*n*,2*nf*)$^1$ and $^{233}$U(*n*,3*nf*)$^1$ respectively, blue solid line, dotted and dashed lines correspond to $^{233}$U(*n,2nf*)$^2$ , $^{233}$U(*n*,3*nf*)$^2$ and $^{233}$U(*n*,3*nf*)$^3$ respectively, dash-dotted and double-dotted dashed lines correspond to PFNS $\langle E \rangle$ with the ε ranges of 0.01—10 and 0—20 MeV respectively.

PFNS ratios $^{239}$Pu($n_{th}$,$f$)/$^{233}$U($n_{th}$,$f$) and $^{233}$U($n_{th}$,$f$)/$^{235}$U($n_{th}$,$f$) of [7, 9, 10]. PFNS of JENDL–4.0 [18] contradict measured PFNS ratio of $^{233}$U($n$,$f$)/$^{235}$U($n$,$f$) at $E_n$~$E_{th}$. Similar contradiction occurs in case of absolute PFNS values (see Fig. 5 and Fig. 6). Calculated PFNS at $E_n$ ~0.5 MeV are compared with the measured PFNS data of $^{233}$U($n$,$f$), $^{235}$U($n$,$f$) [12, 21–23] and $^{239}$Pu($n$,$f$) [7, 14, 22]. It might be concluded that the hardest prompt fission neutrons are emitted in $^{239}$Pu($n_{th}$,$f$) reaction, while the softest PFNS are those of $^{235}$U($n_{th}$,$f$), PFNS of $^{233}$U($n_{th}$,$f$) occupies intermediate position (see Figs. 3, 4, 5, 6).

At $E_n$~0.5 MeV and $E_n$~1.9 MeV calculated/evaluated PFNS deviate from the measured data [8] at the PFNS energy $\varepsilon \gtrsim E_n$ (see Fig. 7). Currently, at $E_n$~E$_{th}$ the renormalization of the model parameters after fitting latest data (see below) on the total kinetic energy of $^{233}$U($n$,$F$) fission fragments amounts to rather small changes of PFNS shapes: for $^{239}$Pu($n$,$f$) decrease by ~2–3% at $\varepsilon \lesssim 1$ MeV, for $^{235}$U($n$, $f$) and $^{233}$U($n$,$f$) PFNS shifts by ~1–2% are observed [24]. Pre–fission neutrons, accompanying fission reaction when the incident neutron energy is higher than threshold of the ($n$,$nf$) reaction, $E_{nnf}$, influence shape of the observed differential PFNS $S(\varepsilon, E_n)$, total kinetic energy of fission fragments $E_F^{pre}$ and fission products $E_F^{post}$, prompt fission neutron number $\nu_p(E_n)$, fragment mass distributions and other fission observables.

Double differential prompt fission neutron spectra $S(\varepsilon, E_n, \theta)$ of neutrons emitted at angle $\theta$ relative to the incident neutron beam, is a superposition of exclusive spectra of pre-fission neutrons, ($n$,$nf$)$^1$, ($n$,$2nf$)$^{1,2}$ or ($n$,$3nf$)$^{1,2,3}$, $\frac{d^2\sigma^k_{nxnf}(\varepsilon, E_n, \theta)}{d\varepsilon d\theta}$ ($x$=1, 2, 3; $k$=1,…,$x$), and spectra of prompt fission neutrons, emitted by the fission fragments, $S_{A+1-x}(\varepsilon, E_n, \theta)$ of ($n$,$nf$), ($n$,$2nf$), ($n$,$3nf$) reactions:

$$\begin{aligned}
&S(\varepsilon,E_n,\theta) = \tilde{S}_{A+1}(\varepsilon,E_n,\theta) + \tilde{S}_{A}(\varepsilon,E_n,\theta) + \tilde{S}_{A-1}(\varepsilon,E_n,\theta) + \tilde{S}_{A-2}(\varepsilon,E_n,\theta) = \\
&\nu_p^{-1}(E_n,\theta)\cdot\{\ \nu_{p1}(E_n)\cdot\beta_1(E_n,\theta)S_{A+1}(\varepsilon,E_n,\theta) + \nu_{p2}(E_n - \langle E_{nnf}(\theta)\rangle)\beta_2(E_n,\theta)S_A(\varepsilon,E_n,\theta) + \\
&\beta_2(E_n,\theta)\frac{d^2\sigma^1_{nnf}(\varepsilon,E_n,\theta)}{d\varepsilon d\varepsilon} + \nu_{p3}(E_n - B_n^A - \langle E^1_{n2nf}(\theta)\rangle - \langle E^2_{n2nf}(\theta)\rangle)\beta_3(E_n,\theta)S_{A-1}(\varepsilon,E_n,\theta) + \beta_3(E_n,\theta)\times \\
&\left[\frac{d^2\sigma^1_{n2nf}(\varepsilon,E_n,\theta)}{d\varepsilon d\theta} + \frac{d^2\sigma^2_{n2nf}(\varepsilon,E_n,\theta)}{d\varepsilon d\theta}\right] + \nu_{p4}(E_n - B_n^A - B_n^{A-1} - \langle E^1_{n3nf}(\theta)\rangle - \langle E^2_{n3nf}(\theta)\rangle - \langle E^3_{n3nf}(\theta)\rangle)\times \\
&\beta_4(E_n,\theta)S_{A-2}(\varepsilon,E_n,\theta) + \beta_4(E_n,\theta)\left[\frac{d^2\sigma^1_{n3nf}(\varepsilon,E_n,\theta)}{d\varepsilon d\theta} + \frac{d^2\sigma^2_{n3nf}(\varepsilon,E_n,\theta)}{d\varepsilon d\theta} + \frac{d^2\sigma^3_{n3nf}(\varepsilon,E_n,\theta)}{d\varepsilon d\theta}\right]\ \}.
\end{aligned} \quad (2)$$

Angle-integrated PFNS $S(\varepsilon, E_n)$ could be obtained by proper integration of $S(\varepsilon, E_n, \theta)$ over phase space ($\theta$ ,φ). In equation (2) the term $\tilde{S}_{A+1-x}(\varepsilon, E_n, \theta)$ is the lumped contribution of $x$-chance fission to the observed (total) PFNS $S(\varepsilon, E_n, \theta)$ at angle $\theta$, $\langle E^k_{nxnf}(\theta)\rangle$ – average energy of $k$–th exclusive pre-fission neutron of ($n$,$xnf$) reaction with spectrum $\frac{d^2\sigma^k_{nxnf}(\varepsilon, E_n, \theta)}{d\varepsilon d\theta}$, $k \leq x$. Note, that spectra $S(\varepsilon, E_n, \theta)$, $S_{A+1-x}(\varepsilon, E_n, \theta)$ and $\frac{d^2\sigma^k_{nxnf}(\varepsilon, E_n, \theta)}{d\varepsilon d\theta}$ are normalized to unity. Index $x$ denotes the fission chance of $^{234-x}$U nuclides after emission of $x$ pre-fission neutrons, $\beta_x(E_n, \theta) = \sigma_{n,xnf}(E_n, \theta)/\sigma_{n,F}(E_n, \theta)$ – contribution of $x$–th fission chance to the observed fission cross section, $\nu_p(E_n, \theta)$ is the observed average number of prompt fission neutrons at angle $\theta$, $\nu_{px}(E_{nx},\theta)$– observed average number of prompt fission neutrons, emitted by $^{234\text{-}x}$U excited nuclides. Spectra of prompt fission neutrons, emitted from the fragments, $S_{A+2-x}(\varepsilon,E_n,\theta)$, as proposed in [25], were approximated by the sum of two Watt [3] distributions with different temperatures, relevant to the light and heavy fragments, the temperature of light fragment being higher:

$$S_{A+1-x}(\varepsilon,E_n) = 0.5\cdot\sum_{j=1}^{2} W_j(\varepsilon,E_n,T_{xj}(E_n),\alpha) \ , \qquad (3)$$

$$W_j(\varepsilon_n,E_n,T_{xj}(E_n),\alpha) = \frac{2}{\sqrt{\pi}T_{xj}^{3/2}}\sqrt{\varepsilon}\exp\left(-\frac{\varepsilon}{T_{xj}}\right)\exp\left(-\frac{E_{vij}}{T_{xj}}\right)\frac{sh(\sqrt{b_{xj}\varepsilon})}{\sqrt{b_{xj}\varepsilon)}}, \qquad (4)$$

$$b_{xj} = \frac{4E_{vxj}^{0}}{T_{xj}^{2}},\ T_{xj} = k_{xj}\sqrt{E_i^{*}} = k_{xj}\sqrt{E_r - E_{fx}^{pre} + U_x}\,. \qquad (5)$$

Here $T_{xj}$ – "temperatures" of light and heavy fragments ($j=l$, $h$) of ($A+2-x$)–th nuclide fissioning after $x$ pre-fission neutrons emission. In equations (3)–(5) energy of center-of-mass system per nucleon is $E_{vxj}^{0} = \frac{A_{hx}}{A_{lx}A_x}\cdot\alpha\cdot E_{fx}^{pre}$. It is assumed that light and heavy fragments emit the same number of neutrons. Though it is a simplification, it is well-known that calculated observed PFNS are not very much sensitive to the inclusion of mass dependence of $\nu_p(E_n,A_{l(h)})$ for light and heavy fragments instead of lumped average number of prompt fission neutrons $\nu_p(E_n)$ [26]. The ratio of "temperatures" of light and heavy fragments $T_{xl}/T_{xh}$, $r$=1.1215 is a semi-empirical parameter which is assumed to be virtually independent on the ($Z$,$N$) values of fissioning nuclide. Its value fits PFNS of $^{233}$U($n_{th}$,$F$), $^{235}$U($n_{th}$,$F$) and $^{239}$Pu($n_{th}$,$F$), the value of $k_{ij}$ depends on the level density parameter. Value of the parameter α– is the ratio of the fragment kinetic energy TKE at the moment of the neutron emission to the TKE at the full acceleration of the fragments, α=0.860 for $^{234}$U, $^{233}$U, $^{232}$U и $^{231}$U fissionning nuclides.

Average energy $\langle E\rangle$ of PFNS in a laboratory system l.s. equals $\langle E\rangle = \langle\varepsilon\rangle + E_v$, here $\langle\varepsilon\rangle$ – is the average energy of prompt fission neutron (PFN) in c.m., while $E_v$ —Watt distribution parameter, i.e. energy of c.m. system in l.s. per nucleon. Usually it is assumed that neutrons are evaporated mostly from fully accelerated fragment, however, it might be assumed that just upon scission some neutrons might be evaporated as well. The kinetic energy of fission fragments $E_v$ is defined with the value of parameter "$\alpha_1$" (see below), in that way measured PFNS of $^{235}$U($n$,$F$), $^{238}$U($n$,$F$), $^{239}$Pu($n$,$F$) and $^{232}$Th($n$,$F$) were described and interpreted [11, 27–33]. Calculated PFNS of $^{239}$Pu($n$,$F$) and $^{235}$U($n$,$F$) [11] either at $E_n \lesssim E_{n2nf}$ and $E_n \gtrsim E_{n2nf}$, were confirmed by the numerous measured data [12–14].

Various pairs of fission fragments with specific values of $E_{fx}^{pre}$ contribute to the observed PFNS, the $\alpha_1$ parameter would be used as $E_{vij} = \alpha_1 E_{vij}^{0}$, to compensate the major approximation of a "pair of average fragments”, average fission energy and average value of TKE, note that $\alpha_1$=1 if $E_n$<6 MeV, while $\alpha_1$=0.8 if $E_n$>12 MeV, and $\alpha_1$ value varies linearly in between.

Total kinetic energy (TKE) values of $E_F^{pre}$, before neutron emission from the fragments, are a superposition of partial TKE values of nuclides contributing to the observed fission cross section:

$$E_F^{pre}(E_n) = \sum_{x=0} E_{fx}^{pre}(E_{nx}) \cdot \sigma_{n,xnf} / \sigma_{n,F} \quad . \qquad (6)$$

The excitation energy $E_{nx}$ of ($A+1-x$) nuclides, formed after emission of $(n,xnf)^{1,..x}$ neutrons, depends on their average energies $\langle E_{nxnf}^k \rangle$, calculated using the exclusive pre-fission neutron spectra $\frac{d\sigma_{nxnf}^k(\varepsilon, E_n)}{d\varepsilon}$:

$$E_{nx} = E_r - E_{fx}^{pre} + E_n + B_n - \sum_{x=0,1\le k \le x} \left( \langle E_{nxnf}^k \rangle + B_{nx} \right). \qquad (7)$$

Kinetic energy $E_F^{post}$ of fission products, which emerge after emission of pre-fission neutrons, but before the $\beta^-$–decay, equals

$$E_F^{post} \approx E_F^{pre} \left( 1 - \nu_{post} / \left( A + 1 - \nu_{pre} \right) \right). \qquad (8)$$

Similar equation for $E_F^{post}$ was employed in [34] at $E_n < E_{nnf}$. At $E_n$ =20 MeV pre-fission neutron contribution $\nu_{pre}$ amounts to ~0.15$\nu_p$. Average prompt fission neutron number $\nu_p(E_n)$ is calculated as

$$\nu_p(E_n) = \nu_{post} + \nu_{pre} = \sum_{x=1} \nu_{px}(E_{nx}) + \sum_{x=1} (x-1) \cdot \beta_x(E_n) \, . \qquad (9)$$

Weak variations of measured TKE values [35], both $E_F^{pre}$ and $E_F^{post}$, in the vicinity of $^{233}$U($n,xnf$) reaction thresholds, as revealed in [36], are due to the decrease of the excitation energy of ($A+1-x$) fissioning nuclides after emission of $x$ pre-fission neutron. Local maxima in the kinetic energy TKE at $E_n > E_{nnf}$, after pre-fission neutron emission for $Z$–even, $N$–odd target nuclide were observed in $^{235}$U($n,F$) [37–39] and $^{239}$Pu($n,F$) [40–42], in the vicinity of ($n,nf$) and ($n,2nf$) reaction thresholds.

The contribution of $^{233}$U($n,nf$) reaction to the observed fission cross section $^{233}$U($n,F$) is governed by the fission probability of $^{233}$U nuclide, which is well-investigated in the $^{232}$U($n,F$) reaction.

Since $B_n(^{233}U)$ ~$B_f(^{233}U)$, fission probability is rather high, while $^{233}U(n,f)$ first-chance fission cross section $\sigma_{n,f}$ decreases with increase of $E_n$ [43] (see Fig. 8), the fission chances distribution for $^{233}U(n,F)$ is much different from those of heavier $N$-odd target nuclides like $^{235}U$ or $^{237}U$. In case of heavier $N$-even target nuclides $^{234,236,238}U$ such shapes of the first-chance fission cross sections $\sigma_{n,f}$ are not observed.

Fission cross section of $^{233}U(n,F)$ in the energy range $E_n$~0.01÷20 MeV is thoroughly investigated [44–54], however, two groups of data sets of [44–50] and [51–54] are systematically different from each other. Absolute measurements of $\sigma_{n,F}$ [46] and [47, 48] at $E_n$~1.93 MeV and $E_n$~14.7, respectively, support the data measured relative to the $\sigma_{n,F}$ of $^{235}U(n,F)$ [44, 45, 49, 50]. Time-of-flight measurements [52–54] are poorly described as regards the absolute values of fission cross sections of $^{233}U(n,F)$ and $^{235}U(n,F)$, they deviate from the data of the first group [44–50] by ~7%.

Present calculated fission cross section of $^{233}U(n,F)$, as those of [4, 5], is consistent with the data of [44–50] (see Fig. 8). Hauser-Feshbach' formalism modelling of $\sigma_{n,F}$ at $E_n \lesssim 0.6$ MeV is sensitive to the positions of discrete transition states at the higher outer saddle point of $^{234}U$ double-humped fission barrier. When 0.6 MeV$\lesssim E_n \lesssim E_{nnf}$, the $\sigma_{n,F}$ is governed by the density of few-quasi-particle states of $^{234}U$ and $^{233}U$ at saddle and equilibrium deformations, respectively. When $E_n > E_{nnf}$, the value of $\sigma_{n,F}$ is governed by the density of excited states of $^{234}U$, $^{233}U$, $^{232}U$ and $^{231}U$ nuclides, which emerge during sequential pre-fission neutron emission, the first pre-fission neutron might be emitted in a pre-equilibrium/semi-direct fashion. Similar analysis was accomplished in case of $^{235}U(n,F)$ [27] and $^{237}U(n,F)$ [55] reactions.

## 3. Pre-fission $(n,xnf)^{1,\ldots x}$ neutron spectra

First pre-fission $^{233}U(n,nf)^1$ neutron spectra are calculated as

$$\frac{d\sigma^1_{nnf}}{d\varepsilon} = \frac{d\sigma^1_{nnX}(\varepsilon)}{d\varepsilon}\frac{\Gamma^A_f(E_n-\varepsilon)}{\Gamma^A(E_n-\varepsilon)}, \qquad (10)$$

here and if not stated otherwise, upper index signifies the first sequential neutron in a specific reaction. Figure 9 shows values of $\beta_x(E_n) = \sigma_{n,xnf}/\sigma_{n,F}$ —partial contributions of the observed fission cross sections $^{233,235}U(n,F)$. Competition of fission and sequential neutron emission for $^{236(234)-x}U$ nuclides defines exclusive neutron spectra of $(n,xnf)^{1,..x}$ and $(n,xn)^{1,..x}$ reactions. Inclusive first neutron spectra of $^{233,235}U(n,nX)^1$ reactions — $\frac{d\sigma^1_{nnX}(\varepsilon)}{d\varepsilon}$ are defined simultaneously with the consistent description of $^{232}U(n,F)$, $^{233}U(n,F)$, $^{234}U(n,F)$, $^{235}U(n,F)$ and $^{235}U(n,2n)$ reaction cross sections. Important feedback comes from the analysis of neutron emission spectra of $^{235}U$+n interaction [24, 57]. Average energies $\langle E^k_{nxnf}\rangle$ of exclusive neutron spectra of $^{233}U(n,xnf)^{1,..x}$ are shown on Fig. 10. Average energies relevant for the $(n,nf)^1$ and $(n,2nf)^1$ reactions are rather hard, while exclusive neutron spectra of $(n,2nf)^2$ and $(n,3nf)^{1,2,3}$ reactions are of the evaporation shape. In the vicinity of the $^{233}$U($n,nf$) reaction threshold the interplay of the exclusive $^{233}$U$(n,nf)^1$ pre-fission neutron spectra and PFNS of the $^{233}$U($n,nf$) fission reaction fragments provokes lowering of the observed average PFNS energy $\langle E \rangle$. The shapes of the observed PFNS at $E_n$~6–11 MeV distinctly correlate with the observed fission cross section, i.e. with fission probability of the nuclides $^{233}$U, which emerge in ($n,nf$) reactions, and ($n,xn\gamma$) reaction cross sections. With increase of the incident neutron energy $E_n$ exclusive neutron spectra $(n,nf)^1$, ($n,n\gamma$) and $(n,2n)^{1,2}$ define the variation of the relative amplitudes of ($n,nf$) neutrons as dependent on the properties of $A+1$ and $A$ nuclides. Figures 11–15 show calculated $^{233,235}$U($n,F$) and measured PFNS of $^{235}$U($n,F$) in the vicinity of $^{233}$U($n,nf$) and $^{235}$U($n,nf$) reaction thresholds. The partial components of the observed PFNS shown on Figs. 11–15 are the ratios to the Maxwellian spectra with the temperature value $T_m$ equal to ~2/3$\langle E \rangle$. Subtracting calculated PFNS of first chance $^{235}$U($n,f$) reaction from the observed PFNS [12] one gets semi-experimental estimate of the $^{235}$U($n,nf$) reaction contribution to the observed PFNS of $^{235}$U($n,F$), as shown at the bottom part of the figures. Figures 11–15 show partial contributions to the PFNS of $^{233,235}$U($n,f$) and $^{233,235}$U($n,nf$) reactions. Variations of PFNS shape with increase of $E_n$ due to the $^{233}$U($n,nf$) and $^{235}$U($n,nf$) reactions contributions, are rather similar, however, the influence of the $\beta_x(E_n)$ values and relative shifts of ($n,nf$) and ($n,2n$) reactions thresholds are quite important. Figures 11–15 demonstrate also pre-fission neutron contributions $\beta_2(E_n)\nu_p^{-1}(E_n)d\sigma_{nnf}^1/d\varepsilon$ of $^{233}$U$(n,nf)^1$ and $^{235}$U$(n,nf)^1$ reactions. Pre-fission neutrons influence observed PFNS both at $\varepsilon \lesssim \langle E \rangle$ and at $\varepsilon \gtrsim \langle E \rangle$. Relative contributions of $(n,nf)^1$ neutrons to the PFNS of $^{233}$U($n,F$) reaction at $E_n$ ~ 6 MeV and $E_n$ ~ 6.5 MeV are higher and lower, respectively, than in case of $^{235}$U($n,F$) reaction. The total or observed PFNS of $^{233}$U($n,F$) and $^{235}$U($n,F$) reactions are similar, since the increase of the contribution to the PFNS of $^{233}$U($n,nf$) reaction, i.e. neutrons, emitted by the relevant fission fragments, is accompanied by the extraordinary decrease of the first chance fission $^{233}$U($n,f$) reaction contribution to the $^{233}$U($n,F$).

When the ($n,nf$) reaction competes with the ($n,n\gamma$) reaction only, the pre-fission neutron shapes depend on the fissilities of $A$ and $A+1$ nuclides rather weakly. When the ($n,2n\gamma$) reaction channel opens, the ($n,nf$) pre-fission neutron shape turns to be rather sensitive to the influence of the exclusive $(n,2n)^1$ and $(n,2n)^2$ neutron spectra. That happens in case of $^{238}$U($n,F$) reaction. For the reaction $^{233}$U($n,nf$) at $E_n \gtrsim 6$ MeV the competition of $^{233}$U$(n,2n)^{1,2}$ neutrons is important, since the threshold of ($n$,2$n$) reaction $E_{n2n}$ ~ 5.7 MeV is rather low. At $E_n$ ~7 MeV and when ε~0.8 MeV, i.e., neutron energy is close to the peak value of the observed PFNS, the contributions of first- and second-chance fission channels are almost equal, i.e. $\tilde{S}_{233}(\varepsilon,E_n) \sim \tilde{S}_{234}(\varepsilon,E_n)$, while for $^{235}$U($n,F$), $\tilde{S}_{235}(\varepsilon,E_n) \sim 1.1\,\tilde{S}_{236}(\varepsilon,E_n)$ (Fig. 13).

The influence of $^{233}$U$(n,nf)^1$ exclusive pre-fission neutrons on the observed $\langle E \rangle$ of $^{233}$U($n,F$) PFNS is quite strong (see Fig. 10). At $E_n$ ~6 MeV relative amplitudes $\beta_2(E_n)\nu_p^{-1}(E_n)d\sigma_{nnf}^1/d\varepsilon$ are systematically higher than those of $^{235}$U($n,F$) reaction. The highest relative amplitude is observed at $E_n$ ~ 6 MeV for $^{233}$U($n,F$) and at $E_n$ ~ 6.5 MeV in case of $^{235}$U($n,F$) reaction (see Fig. 11 and Fig. 12). In the left lower corner of the Fig. 13 the relative amplitudes $\beta_2(E_n)\nu_p^{-1}(E_n)d\sigma_{nnf}^1/d\varepsilon$ of exclusive pre-fission neutron spectra $^{233}$U($n,nf$)$^1$ and $^{235}$U($n,nf$)$^1$ at $E_n$ ~ 6.0, 6.5 and 7.0 MeV are shown, systematic differences of the shapes due to ($n$,2$n$)$^{1,2}$ neutrons competition are evident.

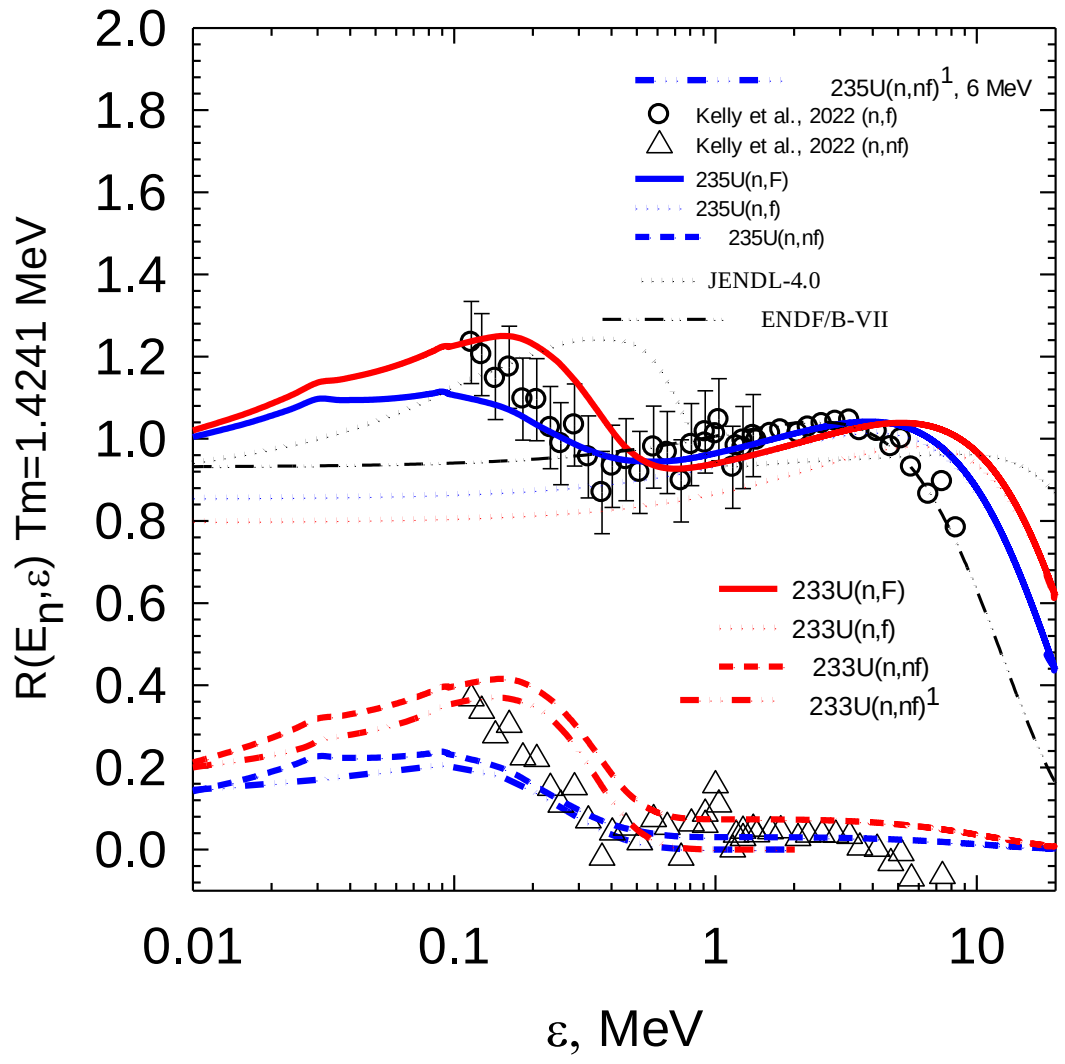

Fig.11. Ratios of partial ($n,xnf$) components of PFNS at $E_n$ = 6,0 MeV of $^{233}$U($n$, $F$) and $^{235}$U($n$, $F$) relative to the Maxwellian-type distribution with $T$ = 1.4241 MeV: the red solid line corresponds to $^{233}$U($n,F$), the red dotted line corresponds to $^{233}$U($n,f$), the red dashed line corresponds to $^{233}$U($n,nf$), the red double-dotted dashed line corresponds to $^{233}$U($n,nf$)$^1$, the blue solid line corresponds to $^{235}$U($n,F$), the blue dotted line corresponds to $^{235}$U($n,f$), the blue dashed line corresponds to $^{235}$U($n,nf$), the blue double-dotted dashed line corresponds to $^{235}$U($n,nf$)$^1$, the white dots correspond to $^{235}$U($n,F$) [12], the white triangles correspond to $^{235}$U($n,nf$) [12], the black double-dotted-dashed line corresponds to the data from ENDF/B-VII [15], the black dotted line corresponds to the data from JENDL-4.0 [18].

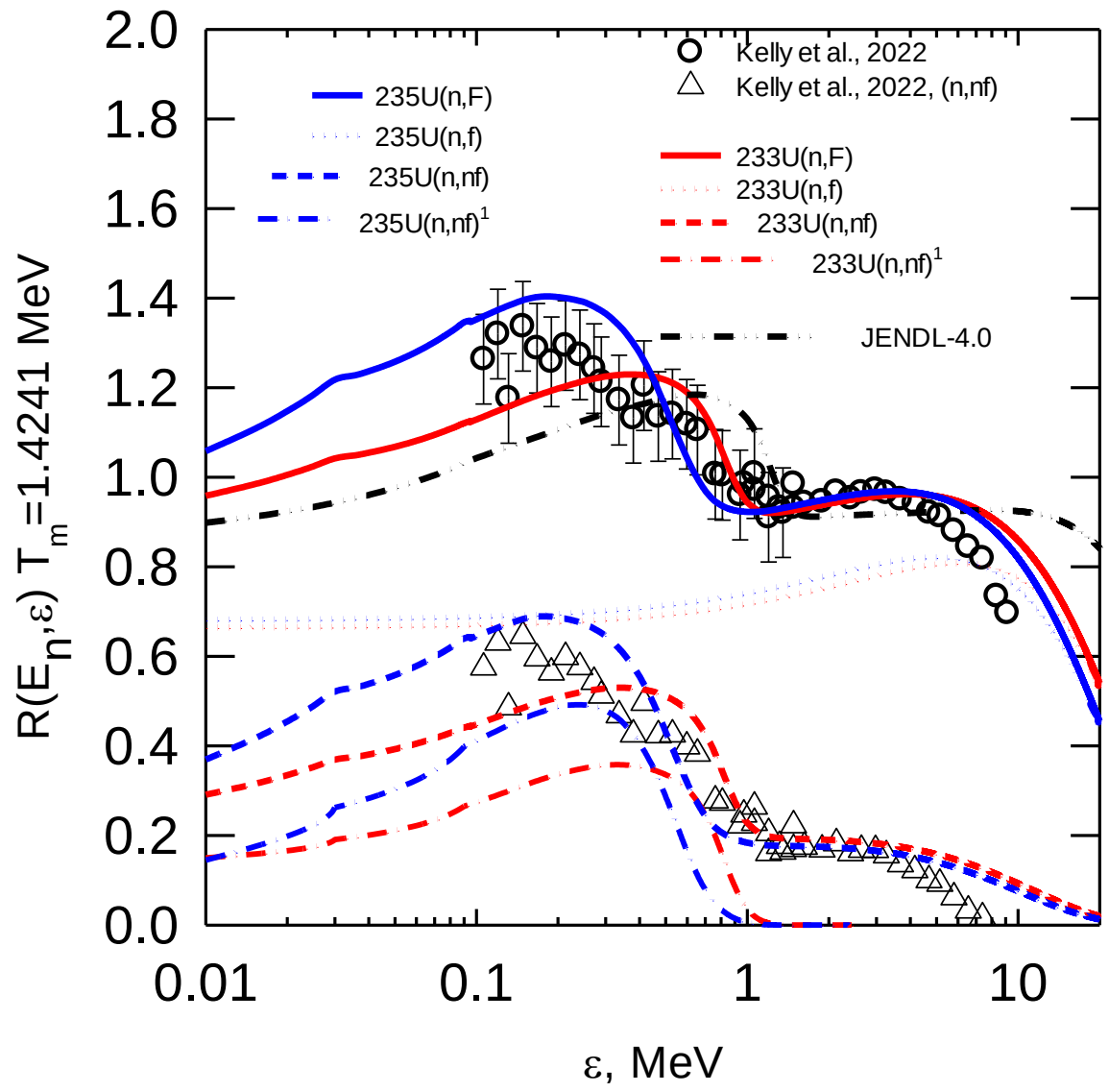

Fig.12. Ratios of partial ($n,xnf$) components of PFNS at $E_n$ = 6.5 MeV of $^{233}$U($n$, $F$) relative to the Maxwellian type distribution with $T$ = 1.4241 MeV: the red solid line corresponds to $^{233}$U($n,F$), the red dotted line corresponds to $^{233}$U($n,f$), the red dashed line corresponds to $^{233}$U($n,nf$), the red double-dotted dashed line corresponds to $^{233}$U($n,nf$)$^1$, the blue solid line corresponds to $^{235}$U($n,F$), the blue dotted line corresponds to $^{235}$U($n,f$), the blue dashed line corresponds to $^{235}$U($n,nf$), the blue double-dotted dashed line corresponds to $^{235}$U($n,nf$)$^1$, the white dots correspond to $^{235}$U($n,F$) [12], the white triangles

correspond to $^{235}$U($n,nf$) [12], the black double-dotted dashed line corresponds to the $^{233}$U($n,F$) data from JENDL-4.0 [18].

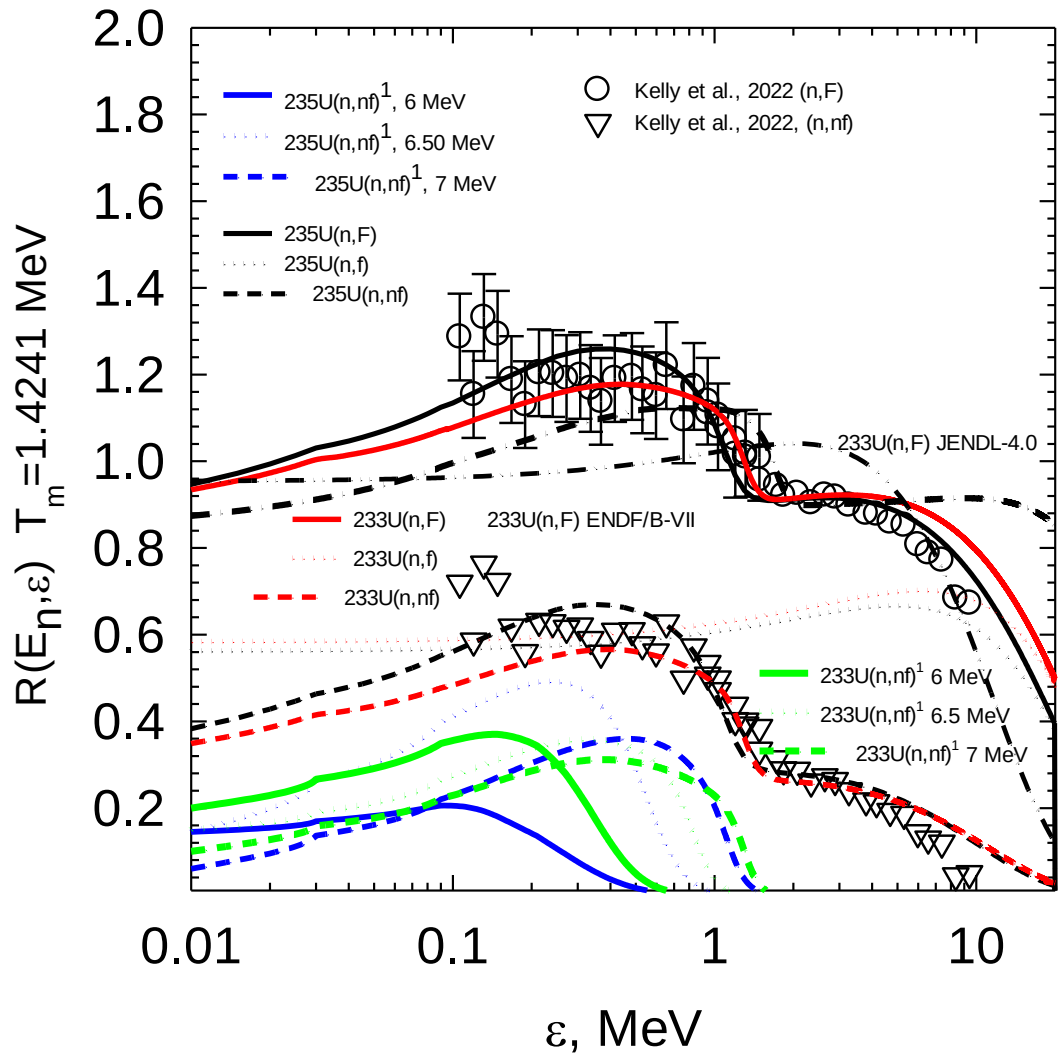

Fig.13. Ratios of partial ($n,xnf$) components of PFNS at $E_n$ = 7,0 MeV of $^{233}$U($n$, $F$) relative to the Maxwellian-_type distribution with $T$ = 1.4241 MeV: the green solid, dotted and dashed lines correspond to the exclusive pre-fission neutron spectra $^{233}$U($n,nf$)$^1$ with the $E_n$ value of 6, 6.5 and 7 MeV respectively, the black solid, dotted and dashed lines correspond to $^{235}$U($n,F$), $^{235}$U($n,f$) and $^{235}$U($n,nf$) respectively, the blue solid, dotted and dashed lines correspond to the exclusive pre-fission neutron spectra of $^{235}$U($n,nf$)$^1$ with the value of 6, 6.5, 7 MeV respectively, the white dots correspond to $^{235}$U($n,F$) [12], the white triangles correspond to $^{235}$U($n,nf$) [12], the black dashdotted line corresponds to the data from JENDL-4.0 [18], the black double-dotted dashed line corresponds to the data from ENDF/B-VII [15].

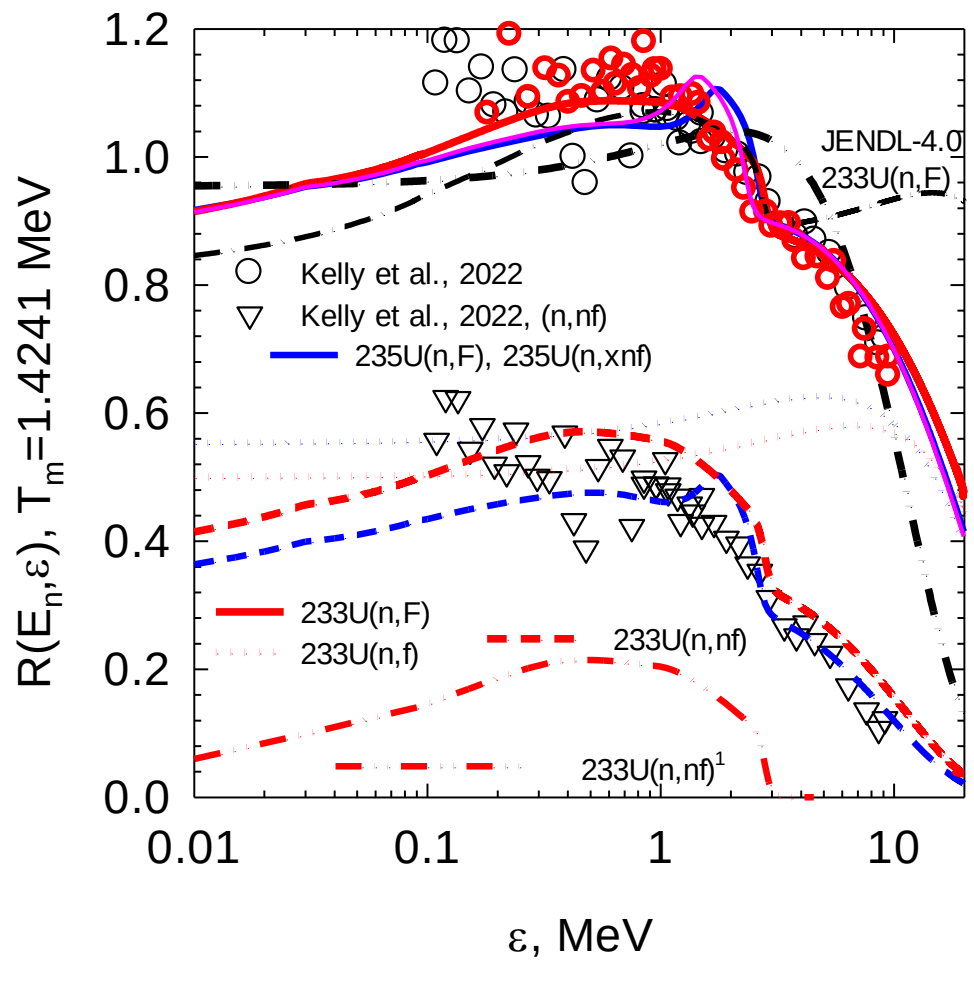

Fig.14. Ratios of partial ($n,xnf$) components of PFNS at $E_n$ = 8.5 MeV of $^{233}$U($n$, $F$) relative to the Maxwellian type distribution with $T$ = 1.4241 MeV: the red solid line corresponds to $^{233}$U($n,F$), the red dotted line corresponds to $^{233}$U($n,f$), the red dashed line corresponds to $^{233}$U($n,nf$), the red double-dotted-dashed line corresponds to $^{233}$U($n,nf$)$^1$, the black dashdotted line corresponds to JENDL-4.0 [18], the black double-dotted dashed line corresponds to the data from ENDF/B-VII [15], the white dots correspond to $^{235}$U($n,F$) [12], the white triangles correspond to $^{235}$U($n,nf$) [12], the red dots correspond to the data of $^{235}$U($n,F$) from [56], pink solid curve corresponds to PFNS at $E_n$ = 8.25 MeV of $^{235}$U($n$, $F$).

Another peculiarity is that the fission fragments of the first chance fission of $^{234}$U nuclides in $^{233}$U($n,f$) reaction are relatively more "heated" than those of $^{233}$U, relevant for the second chance fission $^{233}$U($n,nf$). With increase of the incident neutron energy $E_n$ the partial lumped contributions of $\tilde{S}_{A+1}(\varepsilon, E_n)$ and $\tilde{S}_A(\varepsilon, E_n)$ vary quite appreciably. With the increase of the average energies of exclusive pre-fission neutron spectra $\langle E^k_{nxnf} \rangle$ the contribution of neutrons emitted from the fragments of the second chance fission to the $\tilde{S}_A(\varepsilon, E_n)$, decreases much faster than that of $\tilde{S}_{A+1}(\varepsilon, E_n)$, corresponding to the first chance fission. Comparing the contributions of $\tilde{S}_{234}(\varepsilon, E_n)$ and $\tilde{S}_{233}(\varepsilon, E_n)$ to the observed PFNS of $^{233}$U($n,F$) reaction at $E_{\rm n}$ ~6.0, 6.5, 7.0, 8.5 MeV one may conclude that the fission fragments of $^{234}$U are more "heated", than fission fragments of $^{233}$U (see Figs. 11−14). Measured PFNS data of $^{235}$U($n,F$) at $E_{\rm n}$ ~8.2 MeV [56] available in April of 2025, strongly support the calculated PFNS of $^{235}$U($n,F$) [57]. The calculated PFNS [57] are confirmed by the measured data in terms of the shape and contributions of the $^{235}$U($n,f$) and $^{235}$U($n,nf$) (see Figs. 11−14). The calculated value of the cut-off energy $E_{nnf1}$ of the pre-fission $^{235}$U$(n,nf)^1$ neutrons is also supported by the measured data of [56]. The present modelling of $^{233}$U($n,F$) PFNS use major milestones as in case of $^{235}$U($n,F$) PFNS [57], i.e. PFNS at $E_{\rm n}$ ~$E_{th}$, fission energy balance and fits of $\sigma_{nF}(E_n)$ and $\nu_p(E_n)$ to get exclusive pre-fission neutron spectra $\dfrac{d\sigma^k_{nxnf}(\varepsilon, E_n)}{d\varepsilon}$.

Rather expressive snap-shot is the difference of partial contributions of $^{233}$U($n,F$) and $^{235}$U($n,F$) PFNS at $E_{\rm n}$ ~10.5 MeV (see Fig. 15). Figure 15 demonstrates also ENDF/B-VII evaluation of $^{233}$U($n,F$) PFNS. It seems that in ENDF/B-VII evaluation they almost_waived off the pre-fission neutrons contribution. In case of JENDL-4.0 the contributions of $^{233}$U$(n,nf)^1$ pre-fission neutrons to the observed PFNS is too high as compared with those in case of $^{235}$U($n,F$) reaction [57].

Exclusive pre-fission neutron spectra $\dfrac{d\sigma^k_{nxnf}(\varepsilon, E_n)}{d\varepsilon}$ contain pre-equilibrium/semi-direct components [57–59]. The hard energy part of the exclusive pre-fission neutron spectrum of $(n,nf)^1$ reaction is defined by the fission probability of the excited target nuclide $A$, i.e. $^{233}$U ($^{235}$U in case of $^{235}$U($n,F$) reaction). There are no experimental data on the $^{233}$U+$n$ interaction to estimate more reliably the hard energy part of $^{233}$U$(n,nf)^1$ reaction pre-fission neutron spectrum. To compensate that deficiency one may use the case of $^{238}$U+$n$ interaction, which neutron emission spectra exhibit strong angular anisotropy, when the residual nuclides $^{238}$U remain at excitations of $U$~1÷6 MeV [58,59]. Direct excitation of $^{238}$U ground state band levels $J^\pi = 0^+, 2^+, 4^+, 6^+, 8^+$ was accomplished within rigid rotator model, while that of $\beta$–bands with $K^\pi=0^+$ and $\gamma$–bands with $K^\pi=2^+$, octupole band with $K^\pi=0^-$ was accomplished within soft deformable rotator [60]. The net effect of these procedures is the possibility to get adequate approximation of the angular distributions of $(n,nX)^1$ first neutron inelastic scattering in the continuum for the N-even target nuclides like $^{238}$U or $^{232}$Th and then $^{235}$U, $^{239}$Pu or $^{233}$U targets, as

$$d\sigma^1_{nnX} / d\varepsilon \approx d\tilde{\sigma}^1_{nnX} / d\varepsilon + \sqrt{\frac{\varepsilon}{E_n}} \frac{\langle \omega(\theta) \rangle_\theta}{E_n - \varepsilon} , \qquad (11)$$

here $d\tilde{\sigma}^1_{nnX} / d\varepsilon$ corresponds to the compound emission, which is almost angle-independent with respect to the incident neutron momentum. The adequate approximation of the angular distributions of

$^{238}$U($n,nX$)$^1$ first neutron inelastic scattering in the continuum, which corresponds to $U$~1÷6 MeV excitations for $E_n \lesssim$ 20 MeV was obtained in [58,59]. The approximation of Eq. 1 was employed to describe angular distributions of secondary neutrons in $^{235}$U+$n$ and $^{239}$Pu+$n$ interactions. The anisotropic part of the inclusive first neutron spectrum $^{233}$U($n,nX$)$^1$ relevant for the excitations, comparable with the height of fission barrier of $^{233}$U nuclide, would be evidenced in $(n,xnf)^{1,\dots,x}$ neutron spectra as well. However, it would be pronounced mostly in the neutron spectra of ($n,n\gamma$) reaction [58,59]. That anisotropy is evidenced also in the exclusive spectra of ($n,nf$)$^1$, ($n$,2$nf$)$^1$ and ($n$,2$n$)$^1$ reactions at $E_n$ >12 MeV, their average energies and, consequently. It was first noticed in PFNS of $^{239}$Pu($n,F$) observed at various angles with respect to the incident neutron beam [61].

Inclusive first neutron spectra of ($n,2nX$) reaction, or equivalently, ($n,2nX$)$^1$, depends on the first neutron spectra of ($n,nX$)$^1$ and the probability of neutron emission from the nuclide $A$ as:

$$\frac{d\sigma^1_{n2nX}}{d\varepsilon} = \frac{d\sigma^1_{nnX}(\varepsilon)}{d\varepsilon}\frac{\Gamma^A_n(E_n-\varepsilon)}{\Gamma^A(E_n-\varepsilon)} \qquad (12)$$

Exclusive first neutron spectra of $^{233}$U($n,2nf$) reaction, i.e. $^{233}$U($n,2nf$)$^1$, are represented as:

$$\frac{d\sigma^1_{n2nf}}{d\varepsilon} = \int_0^{E-B_n^{233}} \frac{d\sigma^1_{n2nX}(\varepsilon)}{d\varepsilon}\frac{\Gamma^{232}_f(E_n-B^A_n-\varepsilon-\varepsilon_1)}{\Gamma^{232}(E_n-B^A_n-\varepsilon-\varepsilon_1)}d\varepsilon_1 \qquad (13)$$

Inclusive second neutron spectra of ($n,2nX$) reaction, i.e. ($n,2nX$)$^2$, neutron emission spectra from nuclide $A$, $^{233}$U, are calculated as

$$\frac{d\sigma^2_{n2nX}}{d\varepsilon} = \int_0^{E-B_n^{A}} \frac{d\sigma^1_{X}(\varepsilon)}{d\varepsilon}\frac{\Gamma^{A}_n(E_n-B^A_n-\varepsilon-\varepsilon_1)}{\Gamma^{A}(E_n-B^A_n-\varepsilon-\varepsilon_1)}d\varepsilon_1 \,. \qquad (14)$$

Exclusive spectra of second neutron of ($n,2nf$), i.e. ($n,2nf$)$^2$ are calculated as

$$\frac{d\sigma^2_{n2nf}}{d\varepsilon} = \int_0^{E-B_n} \frac{d\sigma^2_{n2nX}(\varepsilon)}{d\varepsilon}\frac{\Gamma^{A-1}_f(E_n-B^A_n-\varepsilon_1-\varepsilon_2)}{\Gamma^{A-1}(E_n-B^A_n-\varepsilon_1-\varepsilon_2)}d\varepsilon_1 \qquad (15)$$

Hard part of exclusive first neutron spectra $\frac{d\sigma^1_{n2nf}}{d\varepsilon}$ of $^{233}$U($n,2nf$)$^1$ or $^{235}$U($n,2nf$)$^1$ reactions is defined by the fission probability of $^{233}$U or $^{235}$U nuclide, respectively. Exclusive spectra of first and second neutrons of $^{233}$U($n$,$2nf$)$^{1,2}$ or $^{235}$U($n$,$2nf$)$^{1,2}$ are defined by the fission probability of $^{233,232}$U or $^{235,234}$U) nuclides, respectively.

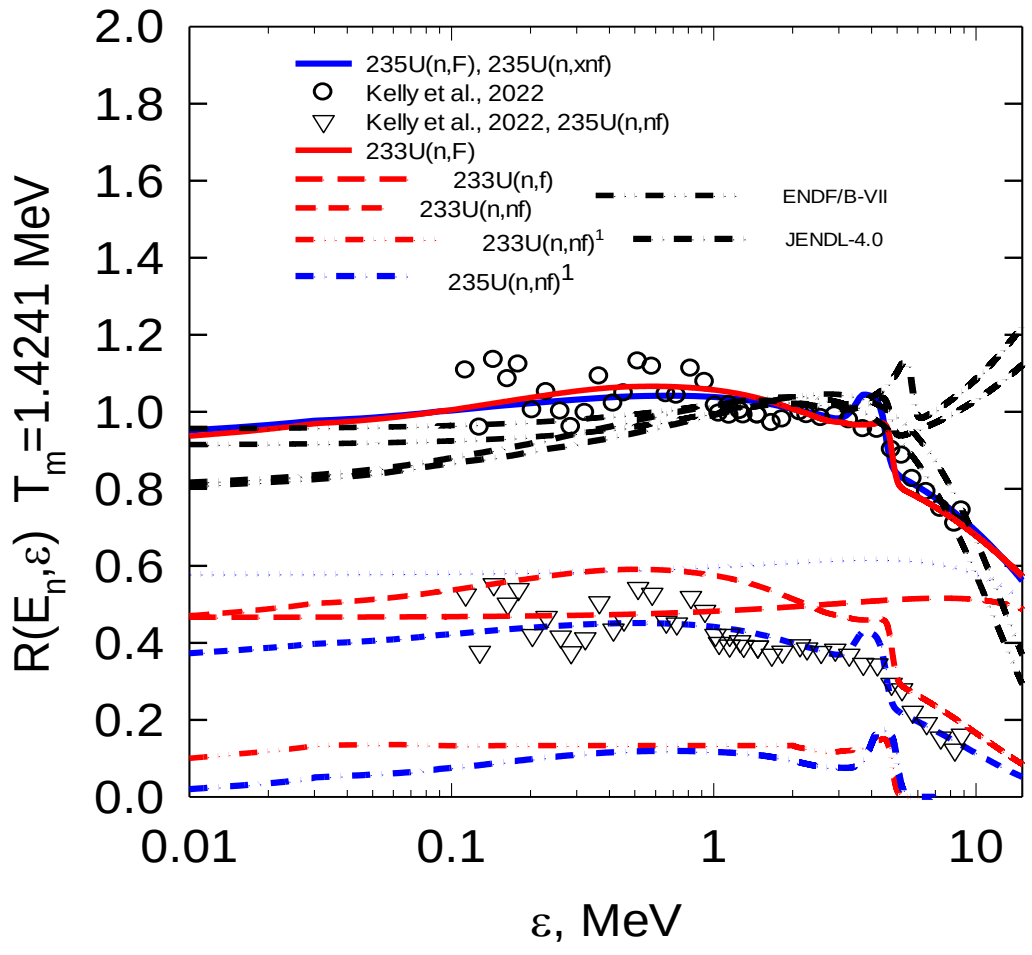

Fig.15. Ratios of partial ($n$,$xnf$) components of PFNS at $E_n$ = 10.5 MeV of $^{233}$U($n$, $F$) relative to the Maxwellian-type distribution with $T$ = 1.4241 MeV: the red solid line corresponds to $^{233}$U($n$,$F$), the red loosely packed dashed line corresponds to $^{233}$U($n$,$f$), the red densely packed dashed line corresponds to $^{233}$U($n$,$nf$), the red double-dotted dashed line corresponds to $^{233}$U($n$,$nf$)$^1$, the blue solid line corresponds to $^{235}$U($n$,$F$), the blue loosely packed dashed line corresponds to $^{235}$U($n$,$f$), the blue densely packed dashed line corresponds to $^{235}$U($n$,$nf$), the blue double-dotted dashed line corresponds to $^{235}$U($n$,$nf$)$^1$ at $E_n$ = 10.5 MeV, the white dots correspond to $^{235}$U($n$,$F$) [12], the white triangles correspond to $^{235}$U($n$,$nf$) [12], the black dashdotted lines corresponds to the data from JENDL-4.0 for the $E_n$ values of 10 and 11 MeV [18], the black double-dotted dashed lines corresponds to the data from ENDF/B-VII for the $E_n$ values of 10 and 11 MeV [15].

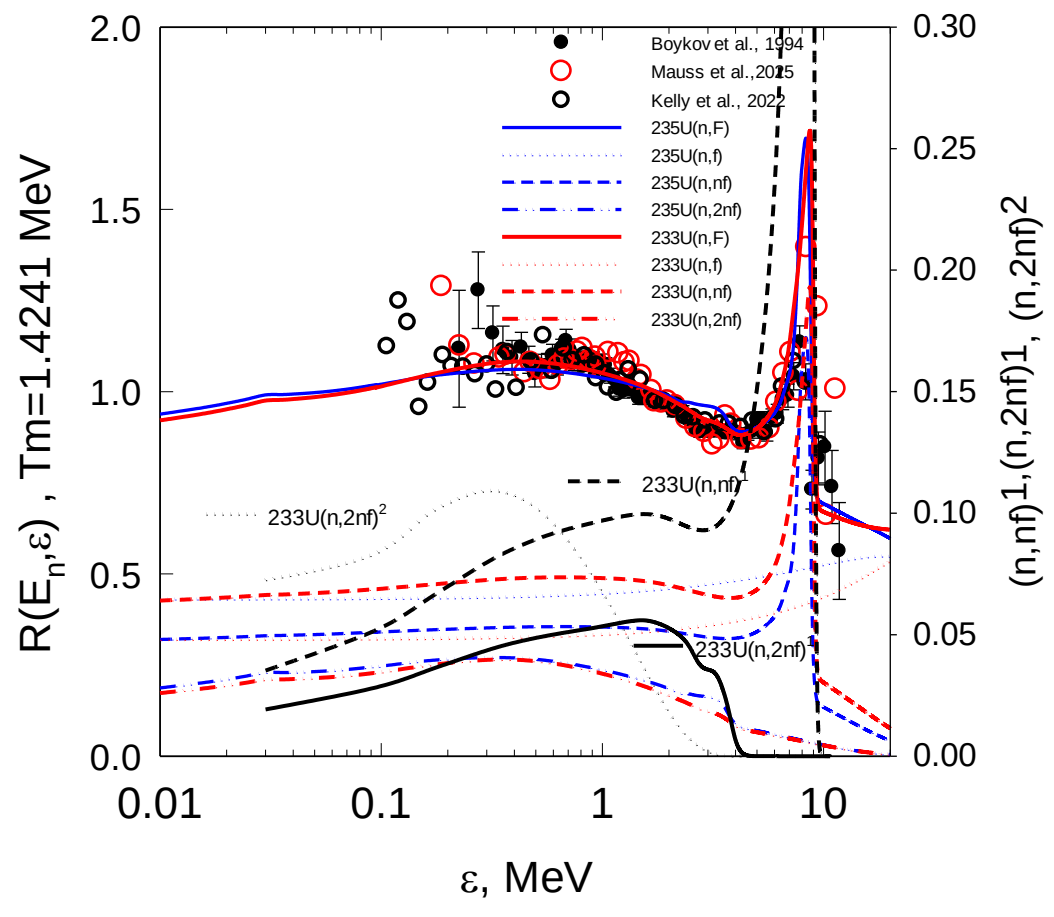

. **

Fig.16. Ratios of partial components of PFNS $E_n$ = 14.7 MeV of $^{233}$U($n$, $F$) relative to the Maxwellian-type distribution with $T$ = 1.4241 MeV: the red solid line corresponds to $^{233}$U($n$,$F$), the red dotted line corresponds to $^{233}$U($n$,$f$), the red dashed line corresponds to $^{233}$U($n$,$nf$), the red double-dotted dashed line corresponds to $^{233}$U($n$,$2nf$), the blue solid line corresponds to $^{235}$U($n$,$F$), the blue dotted line corresponds to $^{235}$U($n$,$f$), the blue dashed line corresponds to $^{235}$U($n$,$nf$), the blue double-dotted dashed line corresponds to $^{235}$U($n$,$2nf$), the black short-dashed line corresponds to $^{233}$U($n$,$nf$)$^1$, the black solid line corresponds to $^{233}$U($n$,$2nf$)$^1$, the black dotted line corresponds to $^{233}$U($n$,$2nf$)$^2$, the white dots correspond

to the data of $^{235}$U($n,F$) from [12], the black dots correspond to the data of $^{235}$U($n,F$) from [62], the red dots correspond to the data of $^{235}$U($n,F$) PFNS from [56].

Figure 16 shows partial contributions to the observed PFNS of $\tilde{S}_{A+1}(\varepsilon,E_n)$, $\tilde{S}_{A}(\varepsilon,E_n)$ and $\tilde{S}_{A-1}(\varepsilon,E_n)$ for $^{233}$U($n,F$) and $^{235}$U($n,F$) reactions as well as measured PFNS data of $^{235}$U($n,F$) [12] and data [62] at $E_n$ ~14.7 MeV. Contribution of the $^{233}$U($n,f$) first chance reaction $\tilde{S}_{A+1}(\varepsilon,E_n)$ to the observed PFNS of $^{233}$U($n,F$) is appreciably lower than in case of $^{235}$U($n,F$) reaction: $\tilde{S}_{234}(\varepsilon,E_n)$~0.6 $\tilde{S}_{236}(\varepsilon,E_n)$. Contributions of the $^{233}$U($n,nf$) and $^{235}$U($n,nf$) reactions $\tilde{S}_{233}(\varepsilon,E_n)$ and $\tilde{S}_{235}(\varepsilon,E_n)$ represent a mirror-like case, in the vicinity of the maximum of $^{233}$U($n,2nf$)$^{1,2}$ neutrons contribution to the PFNS of $^{233}$U($n,F$), one gets $\tilde{S}_{233}(\varepsilon,E_n)$~1.5$\tilde{S}_{235}(\varepsilon,E_n)$. On Fig. 16 are shown partial contributions of exclusive pre-fission neutron spectra $^{233}$U($n,2nf$)$^1$ and $^{233}$U($n,2nf$)$^2$ to the $S(\varepsilon,E_n)$. Note, that lumped contributions of pre-fission and post-fission neutrons of $^{233}$U($n,2nf$) and $^{235}$U($n,2nf$) reactions to the observed PFNS of $^{233}$U($n,F$) and $^{235}$U($n,F$), respectively, almost coincide with each other, i.e. $\tilde{S}_{232}(\varepsilon,E_n)\sim\tilde{S}_{234}(\varepsilon,E_n)$. Contributions to the $S(\varepsilon,E_n)$ of exclusive spectra of pre-fission neutrons $\beta_2(E_n)\nu_p^{-1}(E_n)d\sigma^1_{nnf}/d\varepsilon$, $\beta_3(E_n)\nu_p^{-1}\left[\frac{d\sigma^1_{n2nf}}{d\varepsilon}\right]$ and $\beta_3(E_n)\nu_p^{-1}\left[\frac{d\sigma^2_{n2nf}}{d\varepsilon}\right]$ of $^{233}$U($n,F$) reaction are shown on Fig. 16. The latter corresponding to $^{233}$U($n,2nf$)$^2$ reaction is much softer, than that of $^{233}$U($n,2nf$)$^1$. Exclusive spectra of $^{233}$U($n,nf$)$^1$ and $^{233}$U($n,2nf$)$^{1,2}$ neutrons have intricate dependence on the fission probability of the nuclides $^{234-x}$U, emerging after successive emission of pre-fission neutrons (see Fig. 16).

In the narrow investigated energy range 0.3≲ε≲5 MeV and notwithstanding excessive simplification inherent in Eq. (1), the data [1] and calculated with Eq. (1) PFNS are quite compatible with the calculated PFNS shape for $^{233}$U($n,F$) [4, 5] (see Fig. 17). In case of reaction $^{233}$U($n,F$) [1] average energy of the evaporation (pre-fission) neutron spectrum of Eq. (1) $\langle E_{nxnf}\rangle$=2$T$=1.1 MeV is quite consistent with the average energy of exclusive pre-fission neutron spectra $d\sigma^1_{n2nf}/d\varepsilon$ and $d\sigma^2_{n2nf}/d\varepsilon$ of [4, 5, 11] of present calculations: $\langle E_{n2nf}\rangle = 0.5(\langle E^1_{n2nf}\rangle + \langle E^2_{n2nf}\rangle)$ ~1.2 MeV.

Exclusive first neutron spectra of $^{233}$U($n,2nf$)$^1$, $d\sigma^1_{n2nf}/d\varepsilon$ at $E_n$~14 MeV have rather weak pre-equilibrium component [57,59]. Due to its contribution the average energy of $^{233}$U$(n,nf)^1$ pre-fission neutrons amounts to $\langle E_{nnf}\rangle$ ~3 MeV (see Fig. 10). Present pre-fission neutron spectrum is much different from the evaporation approximation for the pre-fission neutrons used in [1]. Approximation of PFNS with Eq. (1) only qualitatively reproduces the contribution of "soft" pre-fission neutrons, net contribution of pre-fission neutrons with ε≲1 MeV and lumped sum of spectra of neutrons, evaporated from the fission fragments of $^{233}$U($n,f$), $^{233}$U($n,nf$) and $^{233}$U($n,2nf$) reactions (see Fig. 17). Shape of the prompt fission neutrons with energies 5≲ε ≲10 MeV, evaporated from fission fragments, also looks reasonable. Evaluation of ENDF/B–VII [15] adopted, in fact, an evaporation neutron spectrum, either pre-fission or emitted by the fragments. The original ENDF/B–VII PFNS [15] with variable average energy $\langle E \rangle$ and rather arbitrary shape, in the next version of ENDF/B–VIII [17] was abandoned being replaced by the JENDL–4.0 evaluation [18].

The $^{238}$U($n,2nf$)$^1$ and $^{232}$Th($n,2nf$)$^1$ pre–fission neutron signatures at $E_n \gtrsim 15$ MeV were observed in [62] and interpreted in [28, 58, 59]. Due to the poor energy resolution in a double-time-of-flight measurements [12–14, 63] nothing of the kind was observed, but still predicted in exclusive pre-fission neutron spectra. In PFNS data of $^{235}$U($n,F$) [56] some irregularity at $\varepsilon$~3.5 MeV at $E_n \gtrsim 15$ MeV might be attributed to the $^{235}$U($n,2nf$)$^1$ pre-fission neutrons (see Fig. 16), as predicted in [57].

PFNS measurements of $^{233}$U($n,F$) were claimed to be measured by the team accomplished recent measurements of $^{235}$U($n,F$) and $^{239}$Pu($n,F$) PFNS [12–14], which probe incident neutron energy range of $E_n$ ~1.5–20 MeV and prompt fission neutron energy range $\varepsilon$~0.01—10 MeV. The measured PFNS data [12–14] and [56] strongly support the predictions of $^{235}$U($n,F$) PFNS [11, 27, 29, 31, 57] and $^{239}$Pu($n,F$) PFNS [11, 30, 32, 57] up to $E_n$ ~20 MeV. The recent modelling of PFNS for $^{235}$U($n,F$) and $^{239}$Pu($n,F$) reactions [57] might be used to predict the PFNS of $^{233}$U($n,F$). Figures 18 and 19 demonstrate the partial contributions of $\tilde{S}_{A+1}(\varepsilon,E_n)$, $\tilde{S}_{A}(\varepsilon,E_n)$, $\tilde{S}_{A-1}(\varepsilon,E_n)$ and $\tilde{S}_{A-2}(\varepsilon,E_n)$ to the observed(total) PFNS of $^{235}$U($n,F$), $^{239}$Pu($n,F$) and $^{233}$U($n,F$) reactions at $E_n$ =20 MeV. The observed PFNS of $^{235}$U($n,F$) and $^{233}$U($n,F$) almost coincide, as do the partial contributions of $^{233}$U($n,2nf$) — $\tilde{S}_{232}(\varepsilon,E_n)$ and $^{235}$U($n,2nf$) — $\tilde{S}_{234}(\varepsilon,E_n)$. The lumped contribution of neutrons, emitted by the fission fragments of $^{233}$U($n, f$) — $\tilde{S}_{234}(\varepsilon,E_n)$, $^{233}$U($n, nf$) — $\tilde{S}_{233}(\varepsilon,E_n)$ and $^{235}$U($n, f$) — $\tilde{S}_{236}(\varepsilon,E_n)$, $^{235}$U($n, nf$) — $\tilde{S}_{235}(\varepsilon,E_n)$ also almost coincide.

Calculated PFNS of $^{233}$U($n,F$) and $^{239}$Pu($n,F$) are quite different (see Fig. 19), contribution of $^{233}$U($n, f$) — $\tilde{S}_{234}(\varepsilon,E_n)$ in the vicinity of $\varepsilon$~1 MeV is about twice lower than the contribution of $^{239}$Pu($n, f$)— $\tilde{S}_{240}(\varepsilon,E_n)$. Contributions of $^{233}$U($n,nf$) — $\tilde{S}_{233}(\varepsilon,E_n)$ and $^{233}$U($n,2nf$) — $\tilde{S}_{232}(\varepsilon,E_n)$ in the vicinity of PFN energy $\varepsilon$~1 MeV is twice larger than the relevant contributions of $^{239}$Pu($n,nf$) — $\tilde{S}_{239}(\varepsilon,E_n)$ and $^{239}$Pu($n,2nf$) — $\tilde{S}_{238}(\varepsilon,E_n)$ to the $^{239}$Pu($n,F$) PFNS, respectively. These peculiarities are due to the variations of fission probabilities of involved uranium and plutonium nuclides and other fission observables.

## 4. Average prompt fission neutron number

Average prompt fission neutron number $\nu_p(E_n)$ of $^{233}$U($n,F$) was measured in [64–76]. Partial average neutron numbers $\nu_{px}(E_{nx})$ define relative contributions of pre-fission neutrons with spectra $d\sigma^k_{nxnf}/d\varepsilon$ and prompt neutrons, emitted from the fission fragments, $S_{A+2-x}(\varepsilon,E_n)$, see Eq. (2) and equations for $\tilde{S}_{A+2-x}(\varepsilon,E_n)$. To calculate $\nu_p(E_n)$ when $E_n > E_{nnf}$, the data on $\nu_p(E_n)$ for $^{233-x}$U($n,F$) at $E_n < E_{nnf}$ should be used, when available. The partial contributions to the PFNS $S_{A+2-x}(\varepsilon,E_n)$ are correlated with the partial contributions of $\nu_p(E_n)$ and $\sigma_{nF}(E_n)$. Figure 20 compares the calculations with Eq. (9) and measured data $\nu_p(E_n)$. Partial contributions of $^{233}$U($n,f$), $^{233}$U($n,nf$) and $^{233}$U($n,2nf$) reactions are shown, as well as lumped contributions of $^{233}$U($n,nf$)$^1$, $^{233}$U($n,2nf$)$^{1,2}$ and $^{233}$U($n,3nf$)$^{1,2,3}$ reactions, $\nu_{pre}(E_n)$, and post-fission neutrons, emitted from the fission fragments, $\nu_{post}(E_n)$. Partial contribution of $^{233}$U($n,nf$) reaction influences but only weakly smooth energy dependence $\nu_p(E_n)$ around ($n,nf$) reaction threshold $E_{nnf}$. The values of $\nu_{post}(E_n)$ and $\nu_{pre}(E_n)$ of $^{233}$U($n,F$) and $^{235}$U($n,F$)

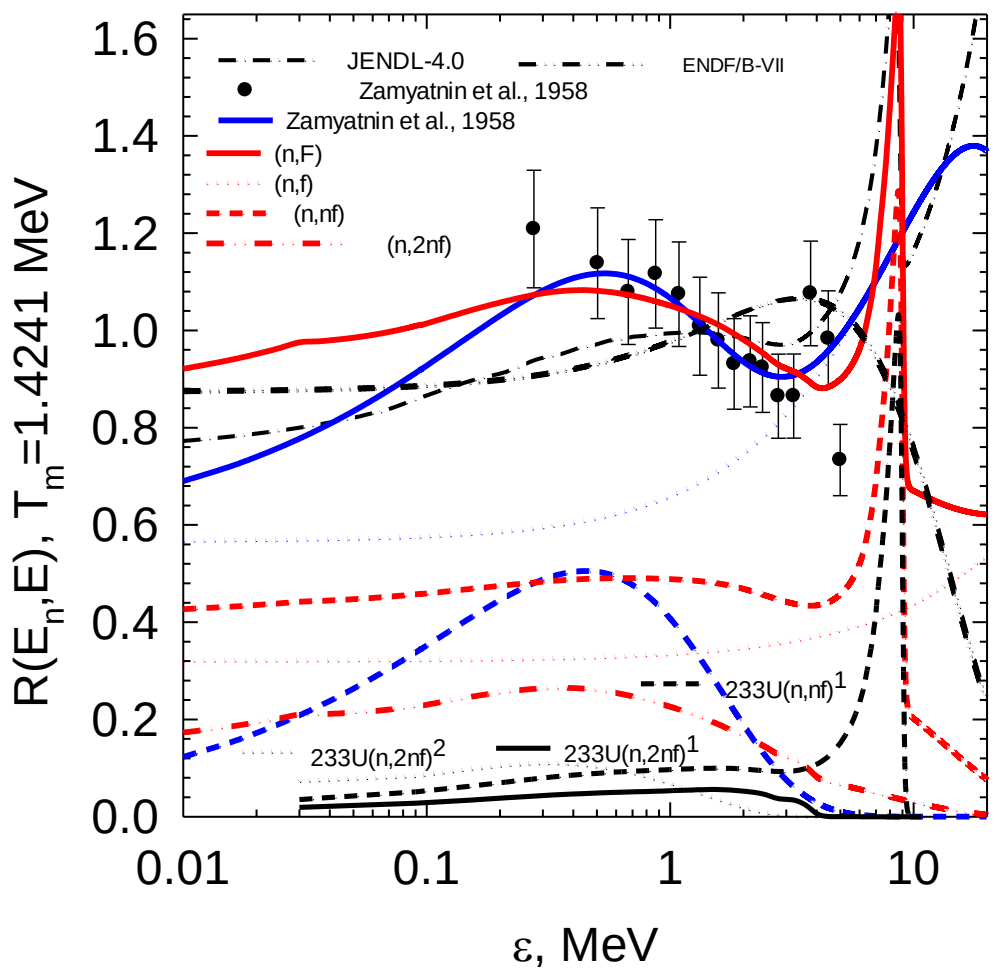

Fig.17. Ratios of partial components of PFNS $E_n$ ~14 MeV of $^{233}$U($n$, $F$) relative to the Maxwellian-type distribution with $T$ = 1.4241 MeV: the red solid line corresponds to $^{233}$U($n,F$), the red dotted line corresponds to $^{233}$U($n,f$), the red dashed line corresponds to $^{233}$U($n,nf$), the red double-dotted dashed line corresponds to $^{233}$U($n,2nf$), the black dashed line corresponds to $^{233}$U($n,nf$)$^1$, the black solid line corresponds to $^{233}$U($n,2nf$)$^1$, the black dotted line corresponds to $^{233}$U($n,2nf$)$^2$, blue solid line correspond to $^{233}$U($n,F$) [1], blue dotted line corresponds to lumped $^{233}$U($n,f$), $^{232}$U($n,f$) and $^{231}$U($n,f$) contributions [1], blue dashed line corresponds to $^{233}$U($n,xnf$)$^{1\ldots x}$ lumped contributions [1], the black dots correspond to the data of $^{233}$U($n,F$) from [1], the black dashdotted line corresponds to the value of $E_n$ = 14 MeV from JENDL-4.0 [18], the double-dotted dashed line corresponds to the value of $E_n$ = 14 MeV from ENDF/B-VII [15].

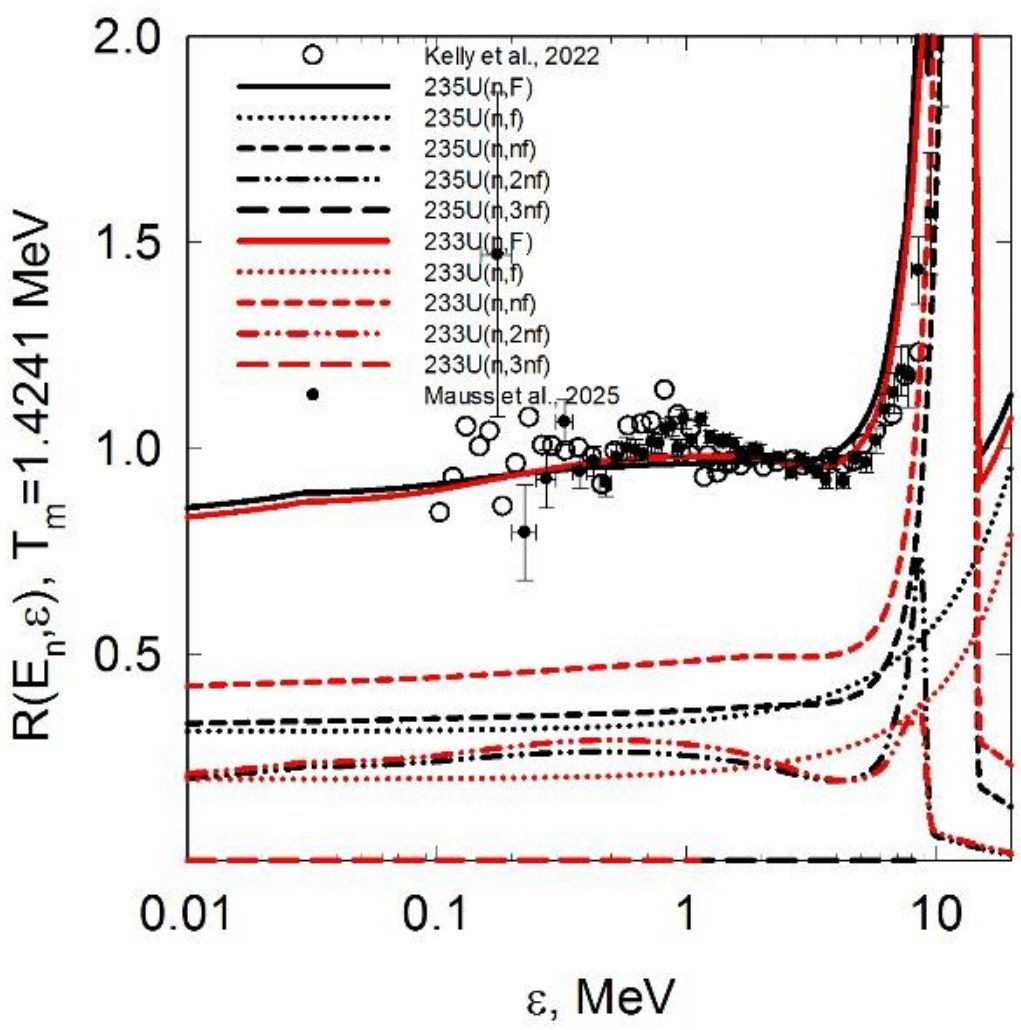

Fig.18. Ratios of partial components of PFNS $E_n$ ~20 MeV of $^{233}$U($n$, $F$) relative to the Maxwellian-type distribution with $T$ = 1.4241 MeV: the red solid line corresponds to $^{233}$U($n$,$F$), the red dotted line corresponds to $^{233}$U($n$,$f$), the red densely packed dashed line corresponds to $^{233}$U($n$,$nf$), the red double-dotted dashed line corresponds to $^{233}$U($n$,2$nf$), the red loosely packed dashed line corresponds to $^{233}$U($n$,3$nf$), the white dots correspond to the data of $^{235}$U($n$,$F$) from [12], the black dots correspond to the data of $^{235}$U($n$,$F$) from [56], the black solid line corresponds to $^{233}$U($n$,$F$), the black dotted line corresponds to $^{235}$U($n$,$f$), the black densely packed dashed line corresponds to $^{235}$U($n$,$nf$), the black double-dotted dashed line corresponds to $^{235}$U($n$,2$nf$), the black loosely packed dashed line corresponds to $^{235}$U($n$,3$nf$).

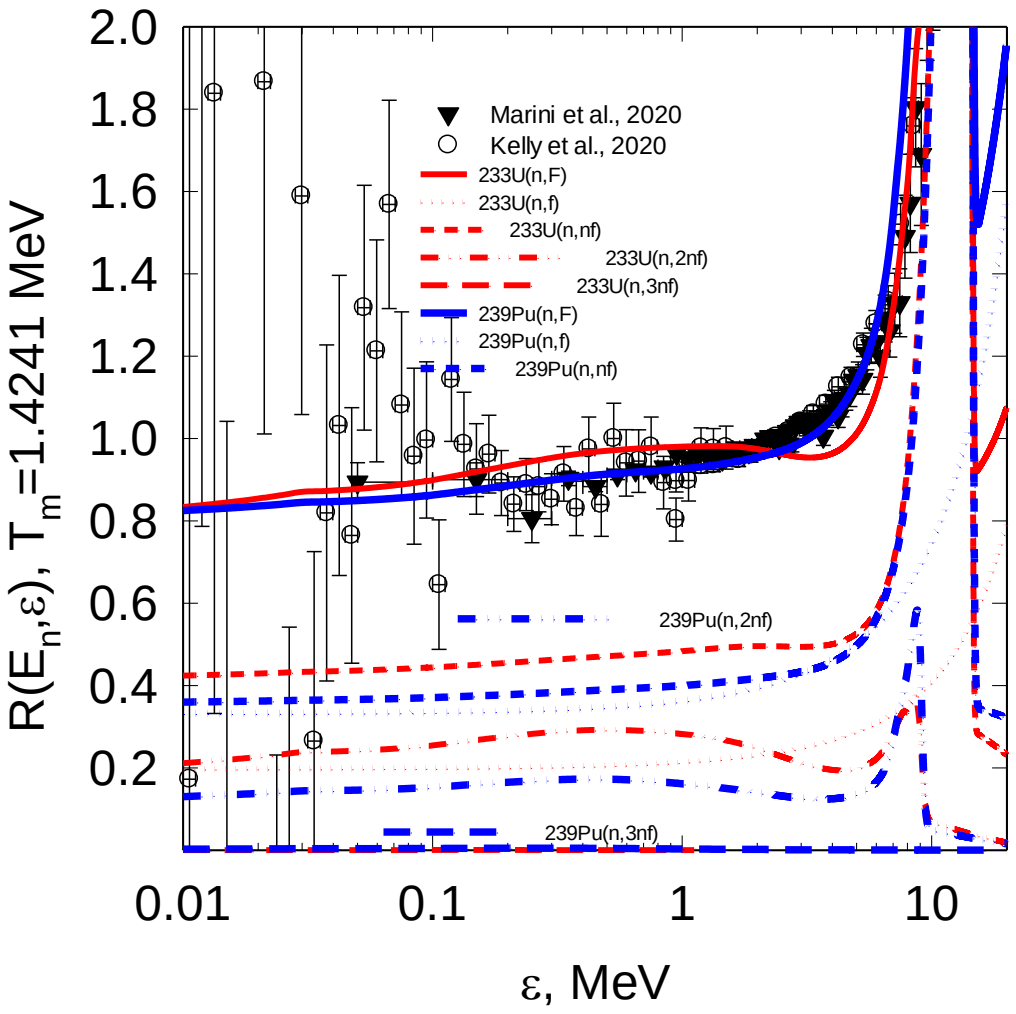

Fig.19. Ratios of partial components of PFNS $E_n$ ~20 MeV of $^{233}$U($n$, $F$) relative to the Maxwellian type distribution with $T$ = 1.4241 MeV: the red solid line corresponds to $^{233}$U($n$,$F$), the red dotted line corresponds to $^{233}$U($n$,$f$), the red densely dashed line corresponds to $^{233}$U($n$,$nf$), the red double-dotted-dashed line corresponds to $^{233}$U($n$,2$nf$), the red loosely dashed line corresponds to $^{233}$U($n$,3$nf$), the blue solid line corresponds to $^{239}$Pu($n$,$F$), the blue dotted line corresponds to $^{239}$Pu($n$,$f$), the blue densely dashed line corresponds to $^{239}$Pu($n$,$nf$), the blue double-dotted-dashed line corresponds to $^{239}$Pu($n$,2$nf$), the blue loosely dashed line corresponds to $^{239}$Pu($n$,3$nf$), the white dots correspond to the data from [13], the black triangles correspond to the data from [14].

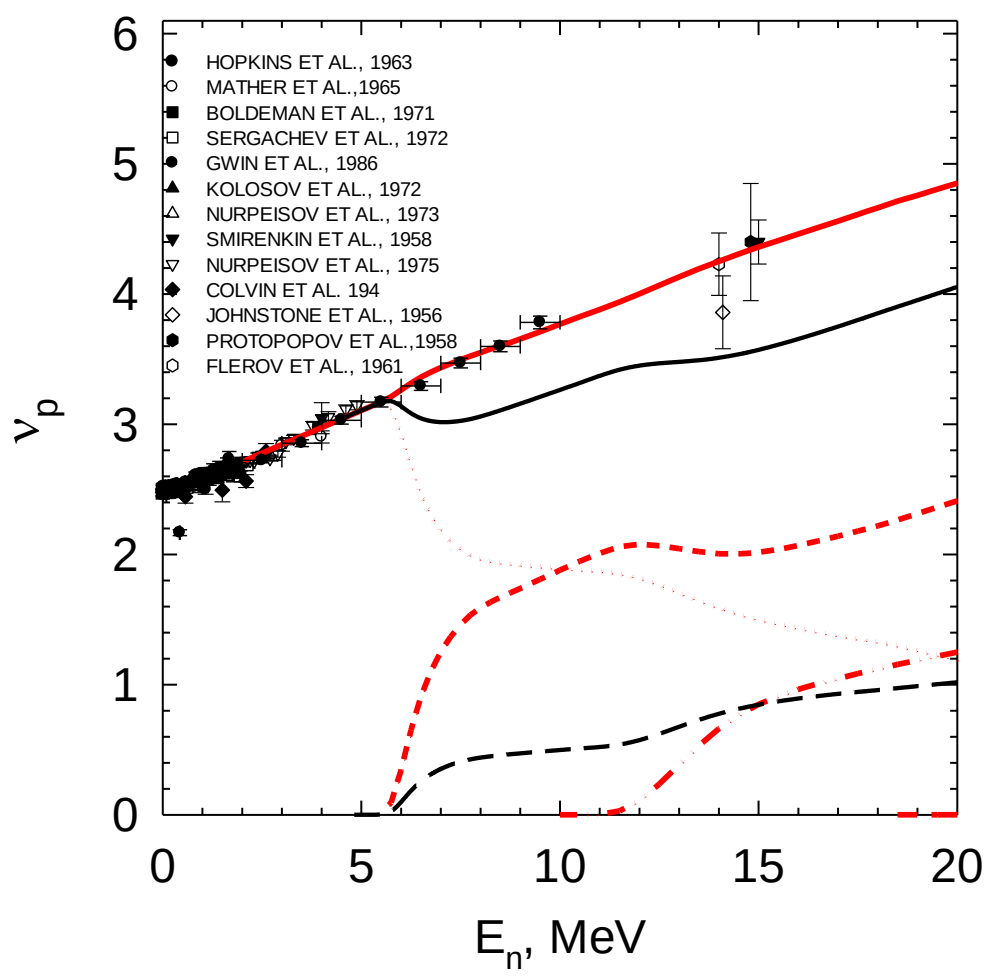

Fig.20. Average number of prompt fission neutrons of $^{233}$U($n$,$F$) and its partial components: the red solid line corresponds to $^{233}$U($n$,$F$), the red dotted line corresponds to $^{233}$U($n$,$f$), the red dashed line corresponds to $^{233}$U($n$,$nf$), the red double-dotted-dashed line corresponds to $^{233}$U($n$,2$nf$), the black solid line corresponds to $\nu_{\rm post}$, the black dashed line corresponds to $\nu_{\rm pre}$. The experimental data are shown by black [64, 65] and white dots [66], black [67] and white squares [70], white up-pointing [68] and down-pointing triangles [69], white [71] and black diamonds [72], black up-pointing [74] and down-pointing triangles [73], white [76] and black hexagon [75].

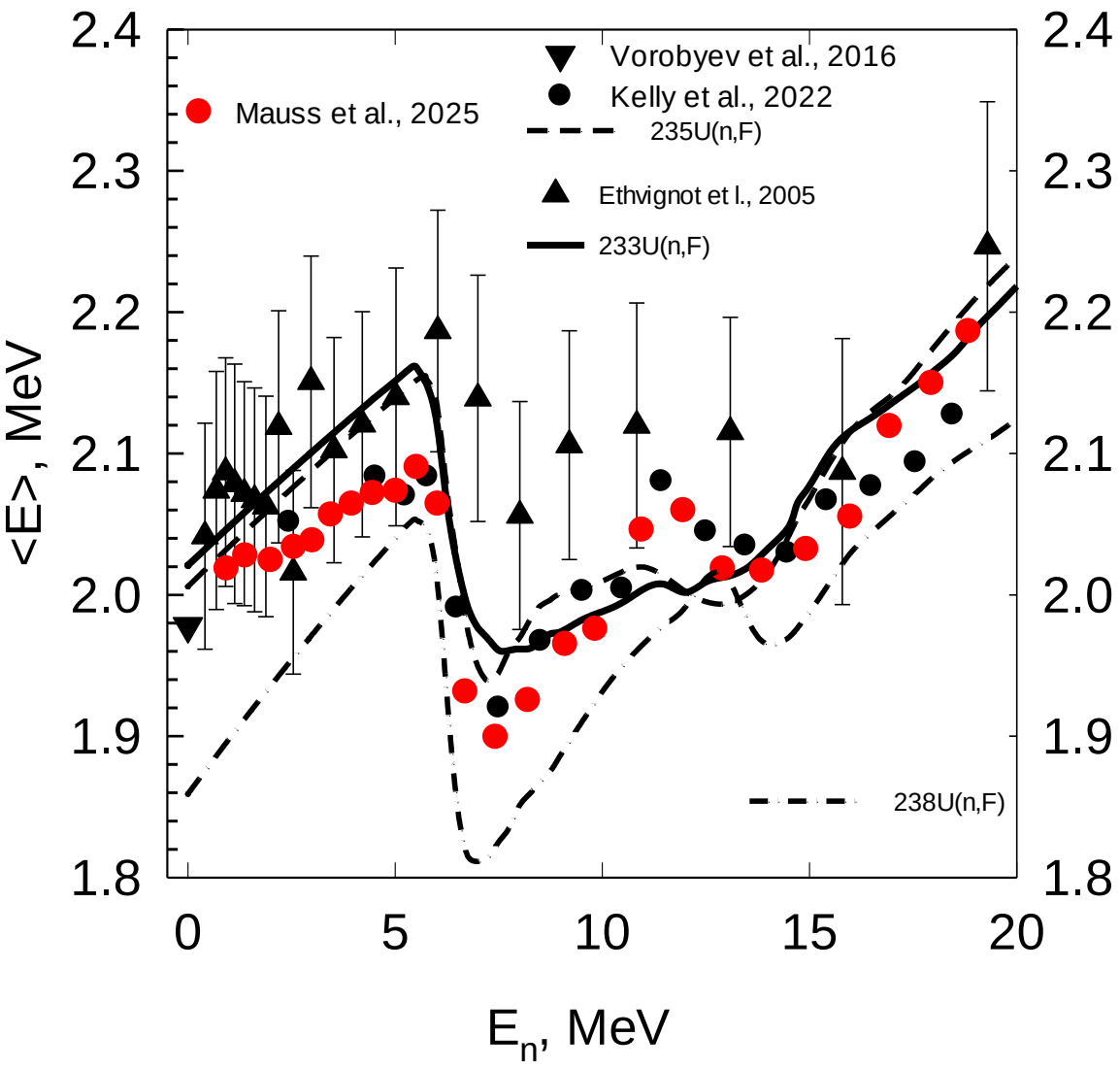

Fig.21. Average energy $\langle E \rangle$ of prompt fission neutrons: the solid line corresponds to $^{233}$U($n$,$F$), the dashed line corresponds to $^{235}$U($n$, $F$), the dashdotted line corresponds to $^{238}$U($n$, $F$) [78], black down-pointing triangles correspond to $^{233}$U($n$,$F$) data from [9], black dots correspond to $^{235}$U($n$,$F$) data from [12], black up-pointing triangles correspond to $^{235}$U($n$,$F$) PFNS data from [83], red dots correspond to $^{235}$U($n$,$F$) PFNS data from [56].

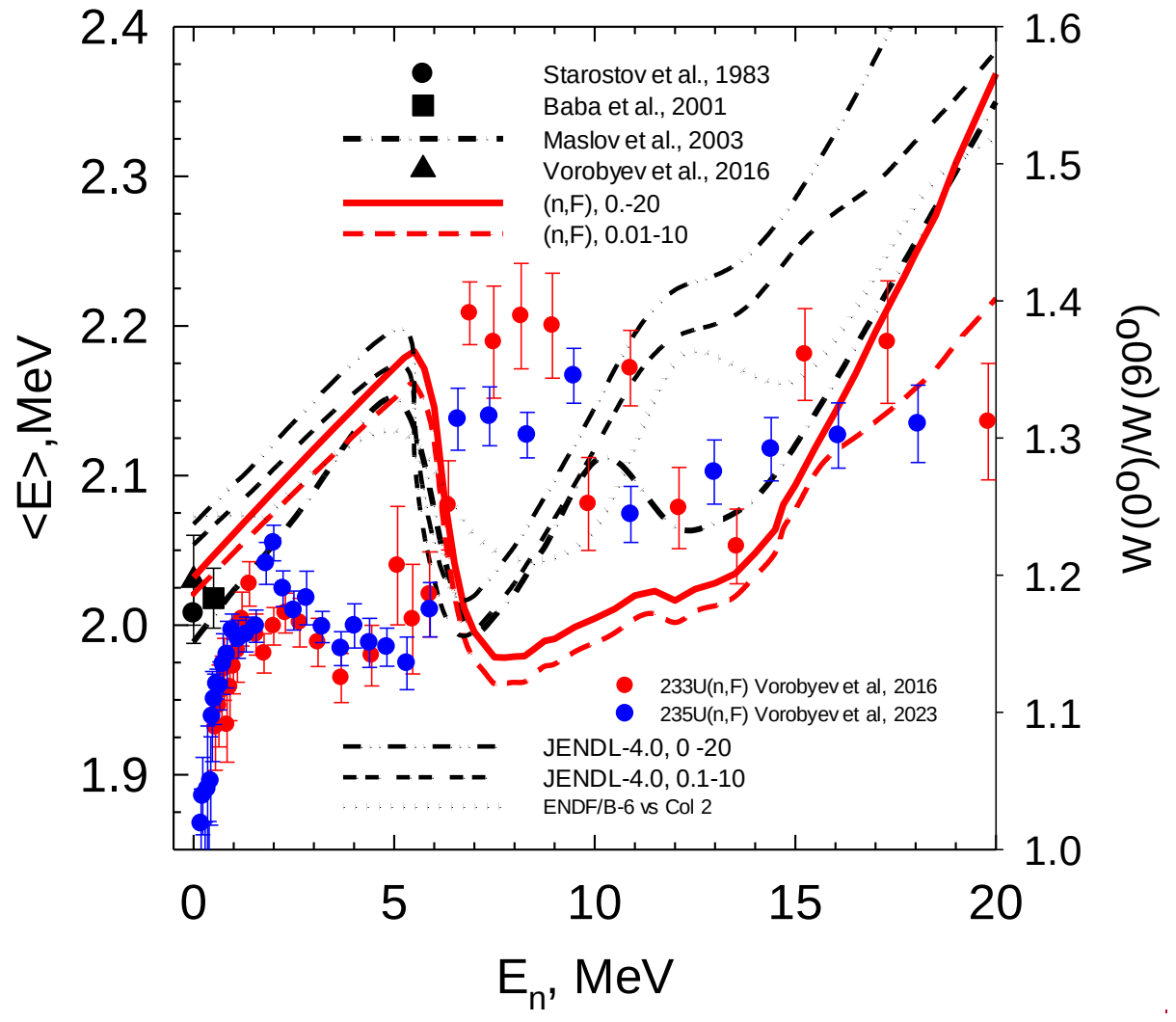

Fig. 22. Average energy $\langle E \rangle$ of prompt fission neutrons of $^{233}$U($n$,$F$): the red solid and red dashed lines are in the ranges of ε from 0 to 20 MeV and from 0.01 to 10 MeV, respectively, the black dashdotted line corresponds to [4, 5], the black dots, squares and triangles correspond to [7], [8], and [9] respectively, the

double-dotted- dashed and short-dashed lines correspond to JENDL-4.0 [18] in the ranges of ε from 0 to 20 MeV and from 0.01 to 10 MeV respectively, black dotted line correspond to ENDF/B-VII [15], red and blue dots correspond to $^{233}$U($n$,$F$) [91 ] and $^{235}$U($n$,$F$) [92] fission fragments anisotropy, respectively.

are influenced by the values of $\beta_x(E_n) = \sigma_{n,xnf}/\sigma_{n,F}$ mostly (see Fig. 9).

## 5. Average energies of PFNS

Average energy of prompt fission neutron spectrum $S(\varepsilon, E_n)$ is its rather rough signature. Figure 21 evidence that the shapes of $\langle E\rangle(E_n)$ in cases of $^{233}$U($n$,$F$) and $^{235}$U($n$,$F$) [24,57] reactions are rather similar. Values of $\langle E\rangle$ [11, 57] for $^{235}$U($n$,$F$) are presented here for the outgoing neutron energy interval $\varepsilon$~0.01–10 MeV. Our estimate of $\langle E\rangle(E_n)$ for $^{235}$U($n$,$F$) [24,57] reproduces the estimate of $\langle E\rangle$, which is based on the measured PFNS data [12], especially around thresholds of $^{235}$U($n$,$nf$) and $^{235}$U($n$,$2nf$) reactions. Estimates of [11, 27, 29, 31] are discrepant with the estimates of [24, 57] in the interval of $E_n$ ~8–10 MeV only.

The present estimate of $\langle E\rangle(E_n)$ for $^{233}$U($n$,$F$) in the interval $\varepsilon$~0–20 MeV is shown on Fig. 22. Though the estimates of PFNS $\langle E\rangle$, given in the data files of ENDF/B–VIII.0 [17] and JENDL–4.0 [18], predict some variation of $\langle E\rangle(E_n)$ for $^{235}$U($n$,$F$) at $E_n$>$E_{nnf}$ , the correlation of these variations with the pre-fission ($n$,$xnf$) neutrons might be considered as arbitrarily imposed. In [17, 18] the correlation of observed PFNS shape with the values $\beta_x(E_n) = \sigma_{n,xnf}/\sigma_{n,F}$ and exclusive $(n,xnf)^{1...x}$ pre-fission neutron shapes is distorted or treated improperly (see Fig. 17). Present estimate of $^{233}$U($n$,$F$) PFNS $\langle E\rangle$ is strictly correlated with the calculated shape of relevant PFNS. The influence of $^{238}$U$(n,nf)^1$ neutrons on $\langle E\rangle$ of $^{238}$U($n$,$F$) PFNS [59] is much stronger than in case of $^{233}$U$(n,nf)^1$ in $^{233}$U ($n$,$F$) reaction. The correlation of $^{238}$U($n$,$F$) PFNS shape with $\langle E\rangle(E_n)$ produces another drop in $\langle E\rangle$ due to the $^{238}$U$(n,2nf)^{1,2}$ neutrons (see Fig. 21). The influence of $^{233}$U$(n,2nf)^{1,2}$ neutrons on $\langle E\rangle$ is much stronger than that of $^{239}$Pu$(n,2nf)^{1,2}$ in $^{239}$Pu($n$,$F$) reaction, while the observed fission cross section values of $^{233}$U($n$,$F$) and $^{239}$Pu($n$,$F$) are quite similar. The correlation of $^{233}$U($n$,$F$) PFNS shape with $\langle E\rangle(E_n)$ produces $\langle E\rangle$ which are quite similar to the $\langle E\rangle$ of $^{235}$U($n$,$F$) PFNS (see Figs. 21, 22). Estimate of the average energy of $^{235}$U($n$,$F$) PFNS [56] for the claimed interval of $\varepsilon$~0–12 MeV, is consistent with the calculations of [57]. One may note, that actually the neutrons were registered in the interval $\varepsilon$~0.18–12 MeV and inherent extrapolations may slightly distort the $\langle E\rangle(E_n)$ values of [56].

## 6. Average total kinetic energies of fission fragments TKE

Local variations of TKE are due to the (*n,xnf*) reactions contributions to the observed fission cross section $\sigma_{n,F}$ and dependence of the partial TKE of $^{234}$U, $^{233}$U, $^{232}$U and $^{231}$U fissioning nuclides on the excitation energy. The contribution of the (*n,nf*) reaction to the observed $\sigma_{n,F}$ of

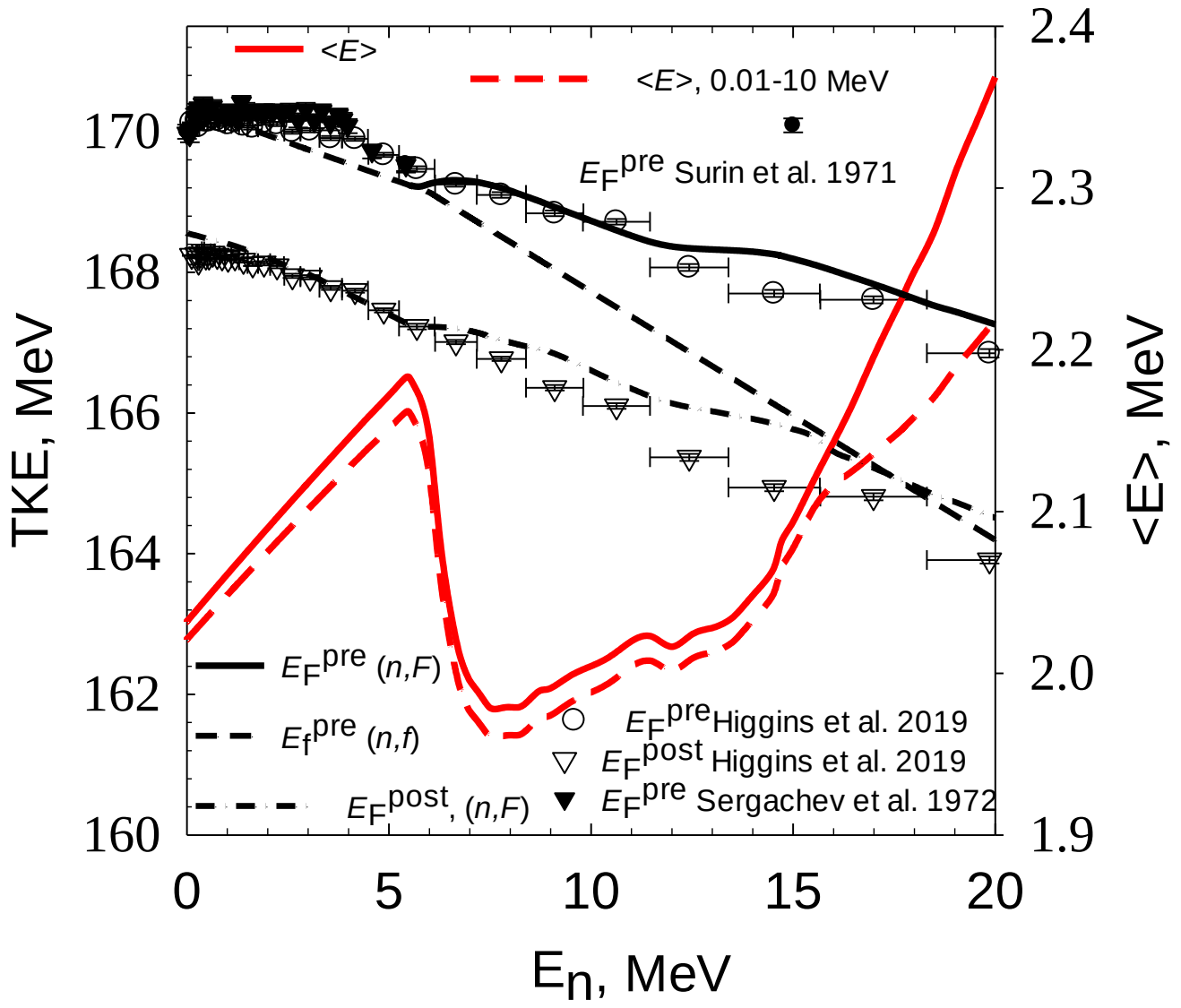

Fig.23. Total kinetic energy (TKE), corresponds to the left vertical axis: the black thick solid line shows $E_F^{pre}$ of $^{233}$U(*n, F*), the dashed line shows $E_f^{pre}$ of $^{233}$U(*n,f*), the dashdotted line shows $E_F^{post}$ of $^{233}$U(*n,F*), the white [35] and black dots [40], and white [35] and black triangles [70] show $E_F^{pre}$ of $^{233}$U(*n,F*). $\langle E \rangle$ values correspond to the right vertical axis: the red solid and dashed lines show $^{233}$U(*n,F*) in the ranges of ε from 0 to 20 MeV and from 0.01 to 10 MeV, respectively.

$^{233}$U(*n,F*) reaction is much higher (see Figs. 8) than in case of $^{235}$U(*n,F*) and $^{239}$Pu(*n,F*) reactions (see Fig. 9), however, the local variations of TKE in $^{233}$U(*n,F*) reaction might be weaker because of rather flat shape of TKE for the first chance fission of neutron-deficient uranium nuclides.

The decrease of $E_f^{pre}$ with the increase of the excitation energy while $E_n < E_{nnf}$, i.e. in the first-chance fission $^{233}$U(*n,f*) (or $^{235}$U(*n,f*)) domain [35], might be attributed to the increase of the distance between fragments at the scission point, as proposed for $^{235}$U(*n,f*) reaction [77]. However, for $^{233}$U(*n,f*) reaction the decrease of TKE with increase of $E_n$ is less evident. Local maxima in TKE at $E_n > E_{nnf}$, due to pre-fission neutron emission, in the vicinity of $^{233}$U(*n,nf*) and $^{233}$U(*n,2nf*) reaction, were observed in [35]. Figure 23 shows that present estimate of TKE in $^{233}$U(*n,F*) reaction is correlated with $\langle E \rangle (E_n)$ of PFNS. The variation of TKE in the vicinity of (*n,xnf*) reaction thresholds due to the decrease of excitation energy after pre-fission neutron emission helps to reproduce measured TKE of [35]. That peculiarity might be considered an indirect proof of calculated shape of PFNS around (*n,xnf*) thresholds, exclusive $(n,xnf)^{1,..x}$ pre-fission neutron spectra and contributions of (*n,xnf*) reactions to the observed fission cross section of $^{233}$U(*n,F*).

Contribution of (*n,nf*) reaction to the $\sigma_{n,F}$ of $^{233}$U(*n,F*), is larger than the calculated contribution of $^{235}$U(*n,nf*) reaction to the observed fission cross section of $^{235}$U(*n,F*), nonetheless the local bumps in TKE around $^{233}$U(*n,2nf*) and $^{233}$U(*n,nf*) reaction thresholds are weaker. That might be due to rather flat dependence on the excitation energy of TKE for first chance fission of $^{232,233,234}$U, contrary to the case of TKE for $^{235,236,237,238,239}$U fissioning nuclides [78, 79]. Even more contrary case is the TKE in $^{232}$Th(*n,F*) reaction [36, 80, 81]. It was observed that TKE of $^{232}$Th(*n,F*) at $E_n < E_{nnf}$ increases with the increase of incident neutron energy $E_n$. That peculiarity explains the occurrence of local minima in TKE in the $^{232}$Th(*n,F*) reaction. To reproduce the observed dependence of $E_F^{pre}$ on $E_n$ in $^{233}$U(*n,F*) reaction one may assume linear dependence of the first-chance fission TKE $- E_{f0}^{pre}(E_n)$.

## 7. Conclusions

A number of peculiarities of fission observables like PFNS, TKE, $\nu_p(E_n)$ or neutron angular anisotropy correlate with the emission of pre-fission (*n,xnf*) neutrons, as predicted for the $^{233}$U(*n,F*) and $^{233}$U(*n,xnf*) and earlier for $^{235}$U(*n,F*), $^{235}$U(*n,xnf*) [67], $^{238}$U(*n,F*), $^{238}$U(*n,xnf*) and $^{232}$Th(*n,F*), $^{232}$Th(*n,xnf*) [59]. We showed that cross ratios of PFNS of $^{233}$U(*n,F*), $^{235}$U(*n,F*) and $^{239}$Pu(*n,F*) reactions are compatible with available measured data [12–14, 82, 84,85]. The correlation of PFNS shape and emissive (*n,xnf*) fission contribution to the observed fission cross section for $^{233}$U(*n,F*) and $^{235}$U(*n,F*) reactions is established. Spectra of $^{233}$U$(n,xnf)^{1,\dots x}$ pre-fission neutrons are rather soft as compared with the spectra of prompt neutrons emitted from the fission fragments. The net effect of these peculiarities is the occurrence of dips in $\langle E \rangle$ in the vicinity of (*n,nf*) and (*n,2nf*) reaction thresholds. Amplitude of the dips in $\langle E \rangle$ of $^{233}$U(*n,F*) PFNS is quite similar to that observed in PFNS of $^{235}$U(*n,F*) reaction, notwithstanding the appreciable differences of $^{233}$U(*n,xnf*) and $^{235}$U(*n,xnf*) reaction contributions to the observed fission cross sections $^{233}$U(*n,F*) and $^{235}$U(*n,F*), respectively. For the reaction $^{233}$U(*n,F*) that might be explained by the relatively large contributions of post-fission neutrons $\nu_{px}(E_{nx})$ as compared with the pre-fission components $\nu_{pre}(E_n)$. In the observed PFNS for the reaction $^{233}$U(*n,F*) the partial components of (*n,f*) and (*n,xnf*) reactions are shown. The partial components of (*n,f*) and (*n,xnf*) reactions are consistent with the $\nu_{pre}(E_n)$. The measurement of PFNS, as described in [12–14], for $^{233}$U(*n,F*) and $^{240}$Pu(*n,F*) at LANSCE beam was first announced in [84]. Before detailed data on PFNS for $^{240}$Pu(*n,F*) [85] become available, the theoretical estimate were presented at [86] and then published in [87]. In fact the data [85] support the semi-blind guess of PFNS for $^{240}$Pu(*n,F*) [86, 87].

Angular anisotropy of prompt fission neutron spectra of nuclides $^{239}$Pu and $^{235}$U is due to the pre-equilibrium emission of $(n,nX)^1$ neutrons some part of which contributes to pre-fission neutrons in (*n,xnf*) reaction [88]. Average energy of $(n,nf)^1$ neutrons depends on the neutron emission angle $\theta$, i.e. fission cross section, prompt neutron number and total kinetic energy depend on $\theta$ as well. Strong sensitivity to forward and backward emission of pre-fission neutrons in $(n,xnf)^1$ reaction was predicted for $^{233}$U($n,F$) [89]. The values of $\beta_2(E_n,\theta\approx 30^o)=\sigma_{n,nf}(E_n,\theta\approx 30^o)/\sigma_{n,F}(E_n,\theta\approx 30)$, shown on Fig. 9, are much larger than those of $\beta_2(E_n,\theta\approx 90^o)$, that explains, at least partly, the anomalous anisotropy predicted for $^{233}$U($n,F$) in [89]. Approximation adopted for $\omega(\theta)$ of $^{239}$Pu($n,F$) and $^{235}$U($n,F$) [57,88] allows to correlate angular anisotropy of $^{233}$U$(n,xnf)^1$ neutrons, relative to the incident neutron momentum, with the emissive fission $^{233}$U($n,xnf$) contributions to the observed fission cross of $^{233}$U($n,F$). The ratios of mean PFNS energies for forward and backward emission of $^{233}$U($n,xnf$) pre-fission neutrons [89] are much larger than relevant ratios for $^{235}$U($n,F$) or $^{239}$Pu($n,F$), but much weaker than in case of $^{238}$U($n,F$) or $^{232}$Th($n,F$) [59,90].

The similarity of dips in average energies $\langle E\rangle$ of $^{233}$U($n,F$) and $^{235}$U($n,F$) PFNS (the depth and obvious dip' asymmetry relative to the largest amplitude) is supported by the fission fragments angular asymmetry with respect to the neutron beam momentum. Figure 22 demonstrates the observed fission fragments angular asymmetry $\mathrm{W(E_n},\theta\approx 0^0)/\mathrm{W(E_n},\theta\approx 90^0)$ (see [93]). The similarity of the wide peaks in $\mathrm{W(E_n},\theta\approx 0^0)/\mathrm{W(E_n},\theta\approx 90^0)$ of $^{233}$U($n,F$) and $^{235}$U($n,F$) reactions (the amplitude and obvious shape asymmetry relative to the peak value) might be explained by the emission of the pre-fission neutrons, which cools down the excited $^{233}$U and $^{235}$U nuclides, fissioning in the second chance fission reaction. Their contribution to the observed $\mathrm{W(E_n},\theta\approx 0^0)/\mathrm{W(E_n},\theta\approx 90^0)$ is responsible for the wide "bumps", shown on Fig. 22.

In case of $^{232}$Th($n,F$), $^{238}$U($n,F$) and $^{240}$Pu($n,F$) reactions much stronger "bumps" in fission fragments angular asymmetry $\mathrm{W(E_n},\theta\approx 0^0)/\mathrm{W(E_n},\theta\approx 90^0)$ were observed by the same PNPI(Gatchina) team [91,92]. The similar correlation with the pre-fission $(n,nf)^1$ and $(n,2nf)^{1,2}$ neutrons as in case of PFNS average energy $\langle E\rangle$ [57,59,86,87] is observed. Note that the amplitude of $\mathrm{W(E_n},\theta\approx 0^0)/\mathrm{W(E_n},\theta\approx 90^0)$ asymmetry which is due to, especially $^{232}$Th($n,nf$) [59,94,95], $^{238}$U($n,nf$) [28,55,59] and $^{240}$Pu($n,nf$) [57,86,87] partial fission channels, is much larger than in case of $^{233}$U($n,nf$) or $^{235}$U($n,nf$) reactions. That is due to the intricate competition of $(n,nf)^1$ and $(n,2nf)^{1,2}$ pre-fission neutrons with the $(n,n\gamma)^1$ and $(n,2n\gamma)^{1,2}$ neutrons [59]. Another important point is the influence on $\mathrm{W(E_n},\theta\approx 0^0)/\mathrm{W(E_n},\theta\approx 90^0)$ of the transition states, i.e. discrete collective states within the pairing gap of the even-even fissioning nuclides $^{232}$Th, $^{238}$U and $^{240}$Pu at saddle point deformations [55].

The measurements of PFNS of $^{233}$U($n,F$) are claimed to be completed in 2025 [96−99] using the experimental setup described in [12, 13] and special neutron detector [99] to cope with α–pile-up and γ-activity of the irradiated sample. Besides evaluations of [4−6, 11, 57], there were a number of efforts to estimate the PFNS of $^{233}$U($n,F$) and $^{235}$U($n,F$) reactions [100−103]. However, in [100−103] they missed the proper influence of exclusive pre-fission neutrons on the observed PFNS and correct estimates of cooling down of residual nuclides $^{236(236)-x}$U after emission of $x$ pre-fission neutrons.

Pre- and post-fission neutron contributions are disentangled here based on the pre-fission neutron spectra prediction and modelling of the spectra of neutrons emitted from the excited fission fragments. That approach produce rather robust prediction of observed PFNS for the $^{233}$U($n,F$) reaction for incident neutron energies $E_n\sim E_{th}\div 20$ MeV with a precision comparable to that attained for the $^{235}$U($n,F$), $^{239}$Pu($n,F$) [57] and $^{240}$Pu($n,F$) PFNS [87, 104].